\documentclass{article}

\usepackage[english]{babel}
\usepackage{authblk}

\usepackage[letterpaper,top=2cm,bottom=2cm,left=3cm,right=3cm,marginparwidth=1.75cm]{geometry}

\usepackage{amsmath}
\usepackage{graphicx}
\usepackage[colorlinks=true, allcolors=blue]{hyperref}
\usepackage{soul}

\usepackage{caption}
\usepackage{subcaption}
\usepackage{adjustbox}
\usepackage{amssymb}
\usepackage{tikz}
\usepackage{rotating}

\title{Statistical treatment of convolutional neural network super-resolution of inland surface wind for subgrid-scale variability quantification}
\author[1]{Daniel Getter}
\author[2,1]{Julie Bessac}
\author[3]{Johann Rudi}
\author[1]{Yan Feng}
\affil[1]{Environmental Science Division, Argonne National Laboratory, Lemont, IL, US}
\affil[2]{Mathematics and Computer Science Division, Argonne National Laboratory, Lemont, IL, US}
\affil[3]{Department of Mathematics, Virginia Tech, Blacksburg, VA, US}
\date{}

\begin{document}
\maketitle
 
\begin{abstract} Machine learning  models have been employed to perform either physics-free data-driven or hybrid dynamical downscaling of climate data. Most of these implementations operate over relatively small downscaling factors because of the challenge of recovering fine-scale information from coarse data. This limits their compatibility with many global climate model outputs, often available between $\sim$50--100 km resolution, to scales of interest such as cloud resolving or urban scales. This study systematically examines the capability of convolutional neural networks (CNNs) to downscale surface wind speed data over land surface from different coarse resolutions (25 km, 48 km, and 100 km resolution) to 3 km. For each downscaling factor, we consider three CNN configurations that generate super-resolved predictions of fine-scale wind speed, which take between 1 to 3 input fields: coarse wind speed, fine-scale topography, and diurnal cycle. In addition to fine-scale wind speeds, probability density function parameters are generated, through which sample wind speeds can be generated accounting for the intrinsic stochasticity of wind speed. For generalizability assessment, CNN models are tested on regions with different topography and climate that are unseen during training. The evaluation of super-resolved predictions focuses on subgrid-scale variability and the recovery of extremes. Models with coarse wind and fine topography as inputs exhibit the best performance compared with other model configurations, operating across the same downscaling factor. 
Our diurnal cycle encoding results in lower out-of-sample generalizability compared with other input configurations. 
\end{abstract}

\maketitle

\section{Introduction}\label{sec:intro}

\subsection{Context and motivations}

Near-surface wind conditions are important for a variety of processes in the Earth system and human activities. Accurate representation of surface wind speed (SWS) in  Earth system modeling has received considerable attention in the past decades, for instance,  in modeling crops \cite{Brisson03}, object drift into the ocean \cite{Ailliot06drift}, and wind-erosion aerosol production \cite{zhang2016quantifying}. In particular, prediction of high winds is critical for assessing extreme weather events and hazards such as storm surges and flooding \cite{de2013wind, westerling2004}. 
Arguably, one of the largest applications of wind predictions is in the wind energy industry, since the wind power density is proportional to the third power of wind speed values near the surface, for example, about 20--50 m above \cite{hennessey1977some}. Wind energy management thus relies heavily on the wind forecasting and uncertainty quantification in the atmospheric boundary layer \cite{Pinson13,Barthelmie2021}. Numerical weather prediction (NWP) or climate models simulate the evolution of temperature, winds, humidity, rainfall, and other atmospheric phenomena based on the mathematical representation of primary dynamic and physical processes in the atmosphere. In these models, the atmosphere is divided into a 3-dimensional grid system made of several millions of grid cells. Each grid cell at any time has unique solutions of the atmospheric variables such as the SWS, and thus offers a means to quantify the wind-related processes near the surface interactively, determined by both large-scale dynamics and local information such as topography and surface fluxes. 

However,  outstanding challenges are associated with the NWP or climate models in predicting the SWS at fine scales in space and/or time. One challenge is the limited model resolution when using modest computational resources; consequently, some of the atmospheric dynamic processes may be under-resolved or resolved insufficiently by those models, leading to uncertainties in the up-scale influences. When using high-resolution meshes and fine timesteps, on the other hand, the computational costs can become prohibitively large. The trade-off between the model predictability and computation motivates the current work. The goal is to reconstruct the spatiotemporal variability in fine-scale SWS based on a combination of machine learning (ML) models and statistical treatment, without the need for large computational resources imposed by running the high-resolution NWP or climate models. We  note that while the present study focuses on the SWS modeling, the  approach can be potentially adapted for resolving other subgrid-scale (SGS) processes and variables such as precipitation and incoming solar radiation. 

Modeling the influence of the subgrid scales on the resolved scales is needed in order to 
accurately represent the entire system \cite{btb15,palmer2019}.
Because a scale separation between resolved and unresolved scales is generally absent with current model resolutions, the upscale influence of SGS processes is not a deterministic function of the resolved state.  
Stochastic models of fine scales have consequently shown great improvement over deterministic models at managing
fine-scale
representations and uncertainties in a number of atmospheric models \cite{grooms2015,dorrestijn2016,berner2017,palmer2019,strommen2019,gagne2020,blein2022}.  
Stochastic subgrid-scale models aim to enhance the missing influence of SGS variability on the resolved state. 
Such stochastic parameterizations, introduced to represent model uncertainty in ensemble forecasts,  have been shown to improve forecast reliability \cite{leutbecher2017} as well as the simulated climate mean state and its variability \cite{christensen2017}. 
 Recent works also have proposed interactive machine-learning treatment of coarse model outputs correcting errors and improving fine-scale representation of some processes such as representation of uncertain small-scale cloud, precipitation, and turbulence processes \cite{bretherton2022,clark2022}. In these works the authors learn nudging tendencies of temperature, humidity, and wind with neural networks and apply these learned corrections online. 

\subsection{Existing work} 
While various super-resolution methods have been utilized in many other applications such as image processing \cite{shi2016real}, downscaling of climate model outputs such as precipitation \cite{wang2021} or sea state \cite{michel2022deep},  only a few works exist on the super-resolution of wind fields and on the corresponding characterization of its subgrid-scale variability. 
Recently, \cite{stengel2020adversarial} developed generative-adversarial models for the super-resolution of climatological wind components and solar fields over the continental United States. 
The study focused on  predictions of high-resolution climate variables using a two-step method of first generating and then learning from intermediate-resolution data, which resulted in a 50x super-resolution total factor. However, it did not focus on  quantifying the associated subgrid-scale variability in those climate variables. 
In addition,  for wind energy purposes \cite{stengel2020adversarial} considered the daily averages of wind components at 100 m height above the surface as input to the learning model. The  approach, however, might not be suitable for 
resolving the diurnal variations, extreme values in the instantaneous SWS, which are more closely coupled with the surface conditions such as the terrain effect than with winds at higher altitudes. For super-resolution of SWS, \cite{kurinchi2021} performed a benchmarking comparison on various deep learning algorithms including a physics-informed resolution-enhancing generative adversarial network; an enhanced super-resolution generative adversarial network (GAN); an enhanced deep super-resolution network; a super-resolution convolutional neural network (SR-CNN)
and cubic interpolation. While the study focused on a relatively small super-resolution factor of 5x on 4-hourly wind components, it highlighted that GAN-based models capture  the spectral dynamics better than CNN-based models do; however, CNN models outperform GANs in terms of visual quality. 

GAN models also pose particular computational issues, which CNNs do not suffer from, because the training of GANs is challenging and often unstable \cite{creswell2018generative}.  The well-known issues during training are convergence difficulties \cite{radford2015unsupervised}, mode collapse of the generative model such that it generates similar samples \cite{salimans2016improved, arora2017generalization}, and vanishing gradients such that the optimization of the generator stalls \cite{arjovsky2017towards}.
As a result, GAN models require careful hyperparameter tuning and selection of model architectures.
Even when overcoming all of these challenges, the interpretability of GAN outputs is an open-ended question, because the generator emits different samples and the evaluation of these sample can lead to different conclusion depending on the evaluation measure \cite{theis2015}.
Because of all of these issues related to GANs, the present work focuses on CNN models.
The advantages of CNNs compared with GANs are that the properties of CNNs are better understood and the training is faster by a few orders of magnitude. CNNs have been successfully leveraged in a wide range of applications, such as, image processing \cite{lecun1999, LeCunBengioHinton15}, image reconstruction \cite{AdlerOktem17}, and inverse problems \cite{FanBohorquezYing19, RudiBessacLenzi22}.

Along with deep learning models, statistical models have been developed to characterize surface wind speed subgrid-scale variability. 
\cite{zhang2016quantifying} proposed to use Weibull distributions, with parameters depending on the resolved mean wind speed, to represent the subgrid-scale variability (15 km and 3 km) within a coarser grid-cell level (around 200 km). 
This method provides a sampling strategy that captures the distribution of fine-scale wind speed within coarser grid cells without accounting for the spatial structure of the subgrid wind fields. 
 \cite{bessac2019,bessac2021scale} investigated, via coarse graining, the difference between ``true'' (at the scale of a high-resolution simulation, 4 km) and ``resolved'  surface wind speed contributing to air-sea fluxes for a range of coarsening scales. These studies considered a short time period (nine days) of a regional convection-permitting model output over the Indo-Pacific Warm Pool region and provided a space-time stochastic model for the subgrid-scale enhancement of unresolved fluxes but without the super-resolution capability.

\subsection{Proposed contributions} 

We propose a convolutional neural network for super-resolution of SWS, called SR-CNN, based on downscaling principles: taking the coarse-gridded surface wind speed (considered as resolved \cite{bessac2019, christensen2020}) and the fine-scale terrain information as inputs, and predicting fine-scale gridded surface wind speed data and its statistical representation.  
Specifically, the study explores  the SWS variability simulated by the Simple Cloud-Resolving Energy Exascale Earth System Model (E3SM) Atmosphere Model (SCREAM) \cite{caldwell2021}, which is a newly released global convection-permitting model developed by the U.S. Department of Energy E3SM project. The new model generates high-resolution outputs at $\sim$3.25 km horizontal grid spacing, capturing many important SGS processes and interaction with Earth's surface unresolved by typical global climate models at $\sim$50--100 km grid spacing. 
The SR-CNN model is constructed to reproduce the fine-scale SWS fields predicted by SCREAM, based on the surface winds gridded at coarse resolutions ($\sim$100, 48, and 25 km) as well as the fine-scale topography. A temporal predictor is implemented in SR-CNN to capture diurnal patterns of SWS. 
Furthermore, we characterize statistically the subgrid-scale variability in the super-resolved wind fields and compare it with an extension of the Weibull parameterization proposed by \cite{zhang2016quantifying} as a mixture of Weibull distributions.
Statistical metrics based on quantiles and spatial correlation are introduced to assess the discrepancies
of the recovered subgrid-scale variability. 
Focusing on winds over the land surface, the SR-CNN is trained with  data over the  southwestern United States and first evaluated with testing data from the same region but unseen during the training. Additionally, more data from two other regions in the eastern Sahara and eastern Asia is used for generalizability evaluation. The selected regions are located in different climate zones with different terrain structure; thus each is representative of the distinctive SWS variability. 
We also use a probabilistic loss function, namely, a Gumbel probability density function, to provide a probabilistic representation of SWS stochasticity at each grid point. This enables sampling and allows for a more flexible representation of extremes compared with the classical loss function based on mean-squared errors.

The study is organized as follows. Section \ref{sec:data} describes the high-resolution numerical model outputs from SCREAM used in the study.  
In Section \ref{sec:methods} we present various configurations of the SR-CNN model that comprises multiple resolution inputs and predictors with adequate loss functions. 
In Section \ref{sec:assessment}  we describe the out-of-sample case studies used to evaluate the proposed model. 
In Section \ref{sec:results} we discuss various statistical characteristics of the super-resolved SWS  and their subgrid-scale variability. In particular, the recovery of extremes in SWS is discussed. 
Section \ref{sec:conclusion} provides some conclusions and perspectives for future work. 

\section{Data}\label{sec:data}

\subsection{SCREAM model outputs}
In this study we use a 40-day simulation of SCREAM from January 20 through March 3, 2020, which is generated as part of the DYnamics of the Atmospheric general circulation Modeled On Non-hydrostatic Domains (DYAMOND) Phase 2 model intercomparison (https://www.esiwace.eu/services/dyamond/winter). High-frequency data outputs at 15-minute intervals are used in the study. As a global convection-permitting version of E3SM, SCREAM is implemented with nonhydrostatic atmosphere dynamics and coupled with the E3SM standard land model and a prescribed ice data model \cite{caldwell2021}. It includes primary atmospheric physics such as parameterizations of cloud microphysics, turbulence, radiation, and a prescribed aerosol implementation, while deep convection is resolved. The horizontal grid for the atmospheric dynamics is a cubed-sphere grid with 1,024 × 1,024 spectral elements on each face, resulting in a grid spacing of $\sim$3.25 km, about 30x finer than the standard E3SM model at $\sim$110 km \cite{golaz2019doe}. 
It also has higher vertical resolution with 128 model vertical layers total, and the vertical grid space in the boundary layer is $\sim$50 m. The high-resolution simulations of SCREAM show significant improvement in the simulated tropical convection, coastal stratocumulus, and precipitation \cite{caldwell2021}. In particular, increased horizontal and vertical resolutions in SCREAM better represent the surface topography and small eddies near the surface, which generate finer-scale spatial and temporal variability in SGS including SWS for the use in super-resolution of coarse-resolution climate model outputs. 
Our region of interest is the $20^{\circ}$ by $20^{\circ}$ region over the continental United States, parts of northern Mexico, and the Gulf of Mexico; see the top left panel of Figure \ref{fig:hourly_wind}. Surface wind speeds exhibit a complex spatial distribution over this region, making it a favorable area to develop the SR-CNN model. We  also use the SCREAM outputs over the other two regions to quantify the SR-CNN model generalizability.

\subsection{Data treatment and initial analysis}
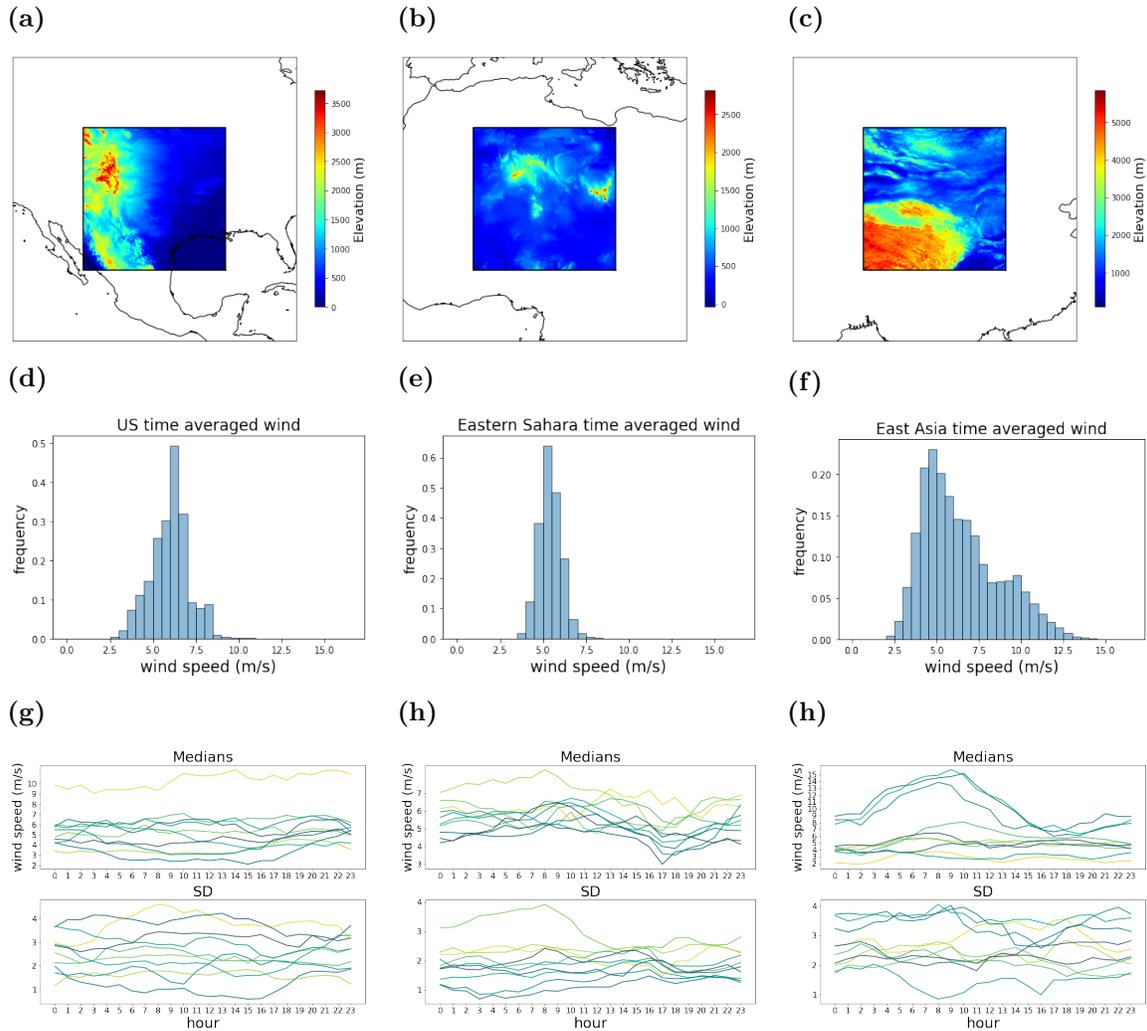
\begin{figure}
     \centering
     \captionsetup[subfigure]{position=above, textfont=normalfont, singlelinecheck=off, justification=raggedright}
     \begin{subfigure}[b]{\textwidth}
         \centering
         \begin{tikzpicture}
         \node (G) 
         {\includegraphics[width=0.31\textwidth]{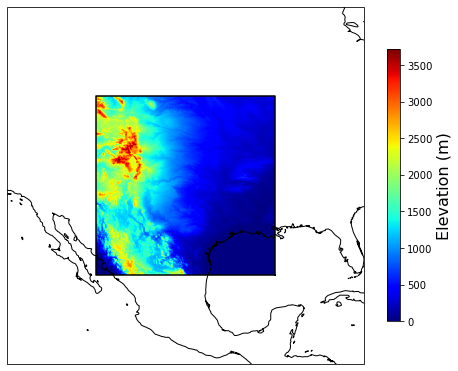}};
         \node [font=\bf, anchor=south west] at (G.north west) {(a)};
         \end{tikzpicture}
         \begin{tikzpicture}
         \node (G) 
         {\includegraphics[width=0.31\textwidth]{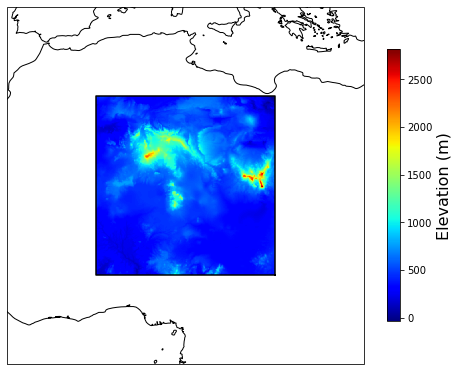}};
         \node [font=\bf, anchor=south west] at (G.north west) {(b)};
         \end{tikzpicture}
         \begin{tikzpicture}
         \node (G) 
         {\includegraphics[width=0.31\textwidth]{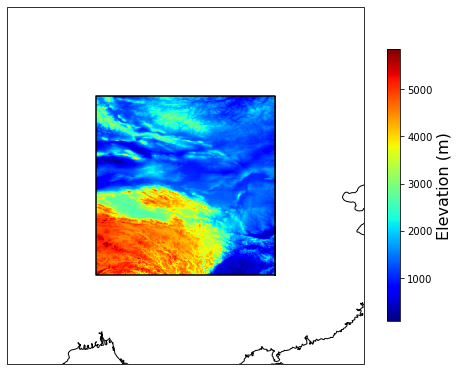}};
         \node [font=\bf, anchor=south west] at (G.north west) {(c)};
         \end{tikzpicture}
     \end{subfigure}
     \begin{subfigure}[b]{\textwidth}
         \centering 
         \begin{tikzpicture}
         \node (G) {\includegraphics[width=.31\textwidth]{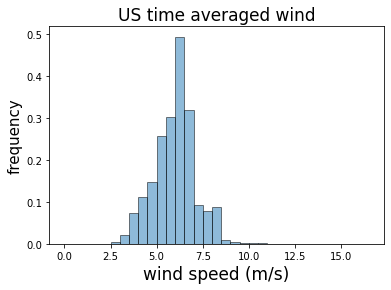}};
         \node [font=\bf,anchor=south west] at (G.north west) {(d)};
         \end{tikzpicture}
         \begin{tikzpicture}
         \node (G) {\includegraphics[width=.31\textwidth]{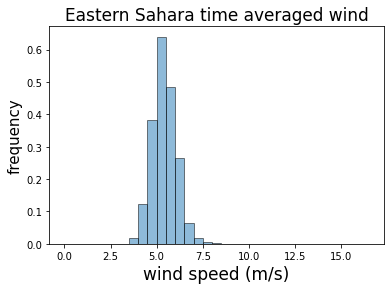}};
         \node [font=\bf,anchor=south west] at (G.north west) {(e)};
         \end{tikzpicture}
         \begin{tikzpicture}
         \node (G) {\includegraphics[width=.31\textwidth]{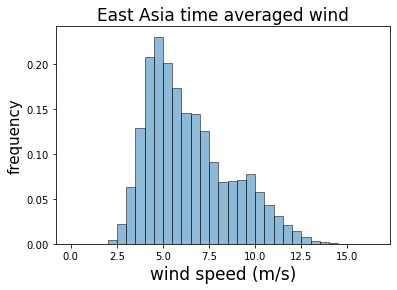}};
         \node [font=\bf,anchor=south west] at (G.north west) {(f)};
         \end{tikzpicture}
     \end{subfigure}
     \begin{subfigure}[b]{\textwidth}
         \centering 
         \begin{tikzpicture}
         \node (G) {\includegraphics[width=.31\textwidth]{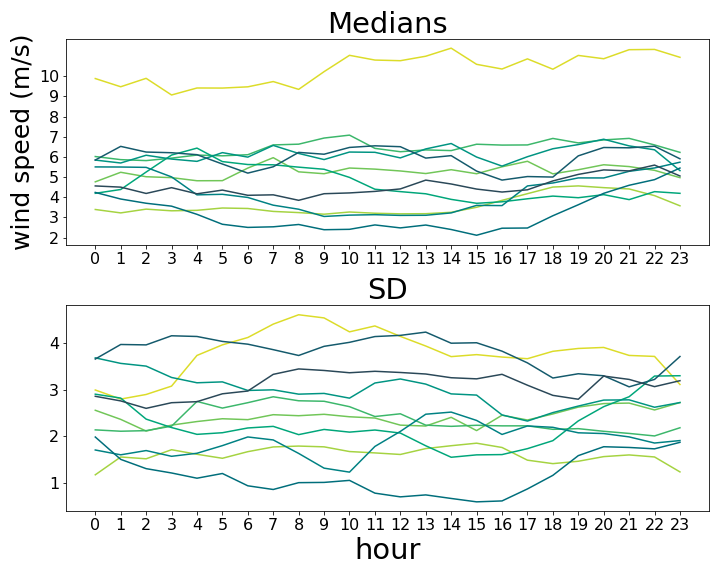}};
         \node [font=\bf,anchor=south west] at (G.north west) {(g)};
         \end{tikzpicture}
         \begin{tikzpicture}
         \node (G) {\includegraphics[width=.31\textwidth]{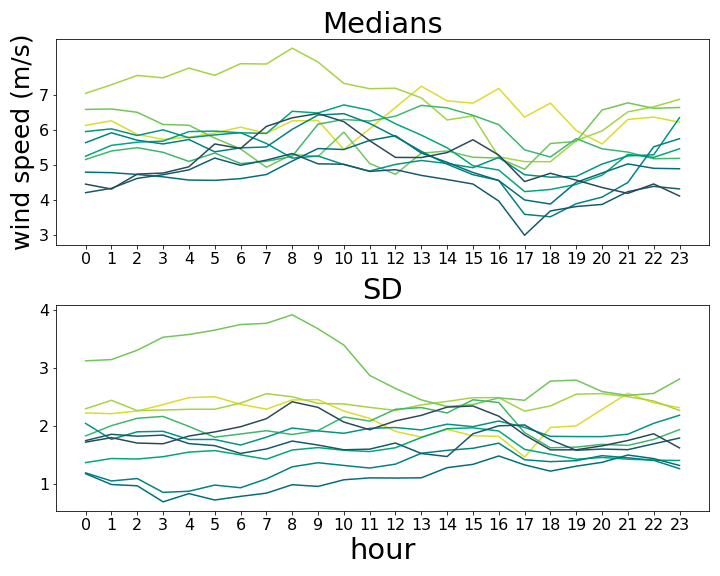}};
         \node [font=\bf,anchor=south west] at (G.north west) {(h)};
         \end{tikzpicture}
         \begin{tikzpicture}
         \node (G) {\includegraphics[width=.31\textwidth]{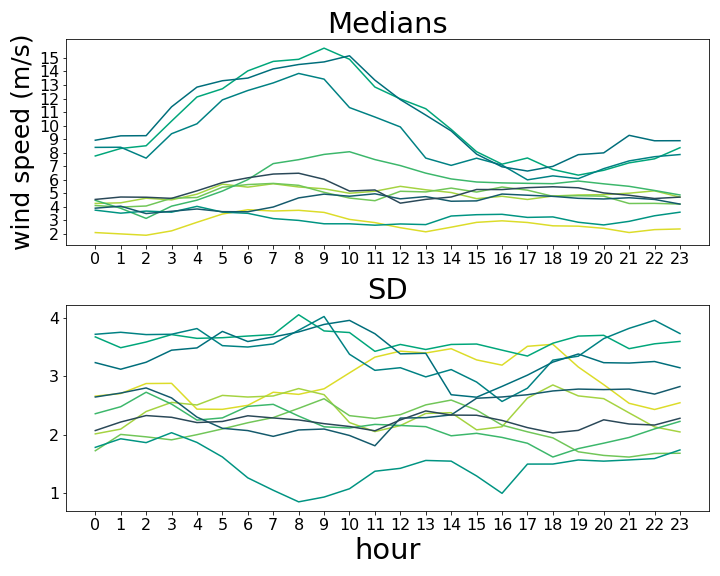}};
         \node [font=\bf,anchor=south west] at (G.north west) {(h)};
         \end{tikzpicture}
     \end{subfigure}
        \caption{Top: Boxed areas are regions used for training (\textbf{a}, US) and out-of-sample testing (\textbf{b}, eastern Sahara and \textbf{c}, eastern Asia), along with associated topographic maps. Central: \textbf{(d-f)} Time-averaged spatial wind speed distribution over each region. Bottom: \textbf{(g-i)} Hourly wind speed distributions from  ten 3 km grid points in each region of interest, across 41 days of data. Note that wind speeds also vary internally by hour within each region. Hours are according to UTC. }
        \label{fig:hourly_wind}
\end{figure}

We incorporate three input variables in our models: surface wind speeds, terrain height (also referred to as topography), and a timestamp
(top central and right panels of Figure \ref{fig:hourly_wind}; all regions are depicted with associated topography).
To emulate coarser climate model simulations, we then perform spatial grid-box averaging over the wind speed data in the region at {\raise.17ex\hbox{$\scriptstyle\sim$}}25 km x 25 km, {\raise.17ex\hbox{$\scriptstyle\sim$}}48 km x 48 km, and {\raise.17ex\hbox{$\scriptstyle\sim$}}100 km x 100 km resolutions  \cite{bessac2019, christensen2020}. 
We additionally interpolate high-resolution terrain height as one of our model predictors. Topography is a known driver of the spatial distribution of surface wind speed \cite{breckling2012}; hence we incorporate this as a way of directly encoding high-resolution information into our model separately from wind speed. 

The central panels of Figure \ref{fig:hourly_wind} describe the spatial distribution of winds averaged over our 41-day time window. Note that each region exhibits unique wind speed variability (see Section 
\ref{sec:results} for further discussion).
The lower panels of Figure \ref{fig:hourly_wind} highlight the different temporal distribution of wind speeds across our 3 regions at ten randomly sampled 3 km grid points. 
We expect to observe various diurnal patterns of wind within each region in link with the local geography, for instance, in link with the distance to the coast.  
Eastern-Asia shows the most pronounced diurnal pattern, with a sharp change in median intensity and variability of wind speed at numerous points. 
We aim to incorporate the diurnal information into the models, and we discuss in the following section a solution. 

\section{Methods}\label{sec:methods}

\subsection{Super-resolution model: SR-CNN}

In this work we construct several super-resolution models at varying resolution gaps and varying number of model inputs. 
We compare the performances of these models with a special  focus on  subgrid-scale variability. We use convolutional neural networks  \cite{lecun1999} to account for local spatial correlations of wind fields, and we look at three differences in resolution between coarse wind speed input and final output:
  25 km to 3 km (25K hereafter) resulting in 8x downscaling factor, 
  48 km to 3 km (48K) resulting in 16x downscaling factor, and
  100 to 3 km (100K) resulting in 32x downscaling factor.
Within each resolution jump, SR-CNN models are created to take between one and three inputs: wind speed; wind speed and topography; or wind speed, topography, and temporal information.
Figure \ref{fig:CNNs} presents example input and output wind speeds at the three different resolutions (25K, 48K, 100K). 
In the following, we describe the three different classes of our model designs, which are distinguished by the models' inputs. Illustrations of these models are depicted in Figure~\ref{fig:CNNs}.

\subsubsection{1-channel input -- coarse wind speed}
This model
is designed to take as its only input coarsened wind speed at one of the three resolutions of interest. This input is passed through a series of convolutional layers with varying filter counts, through which the model learns low- and high-level spatial features within the data.
The output of the final convolutional layer is passed to a depth-to-space operation,which maps the filters of the final convolutional layer to the final resolution of 3 km $\times$ 3 km of the downscaled wind speeds.
For instance, 64 filters of the last convolutional layer in Figure~\ref{fig:CNNs}(a) map one ``pixel'' of coarse wind speeds to 8$\times$8 ``pixels'' of downscaled wind speeds, which yields a downscaling factor of 8x.

\subsubsection{2-channel input -- coarse wind speed, fine-scale topography}
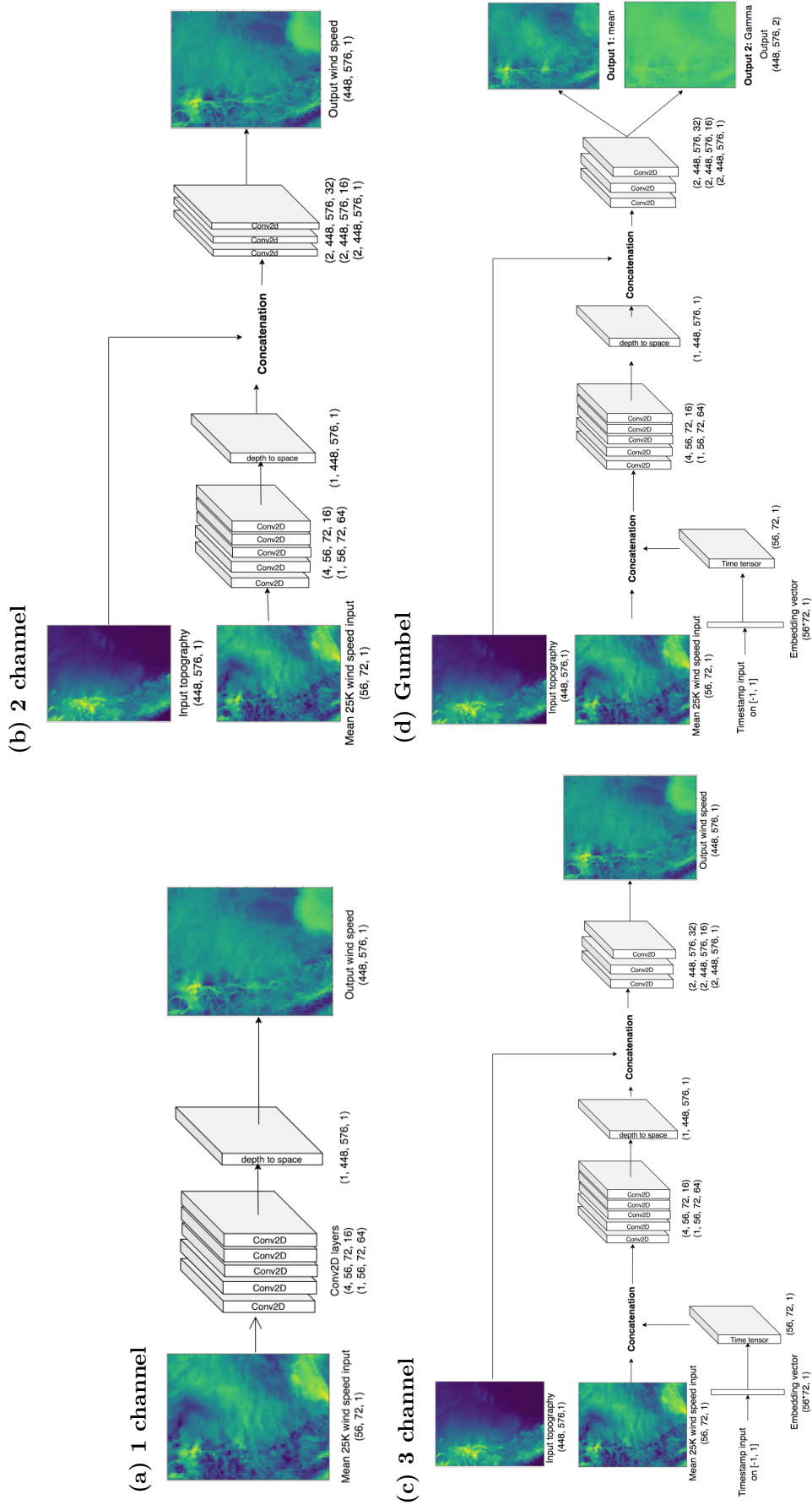
\begin{sidewaysfigure}
    \centering
     \begin{subfigure}[b]{\textwidth}
         \centering
         \begin{tikzpicture}
         \node (G) {\includegraphics[width=.4\textwidth,height=\textheight,keepaspectratio]{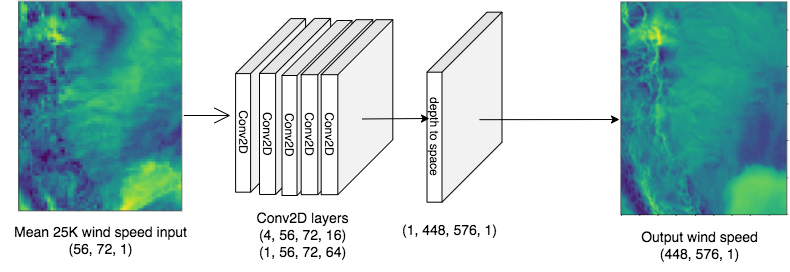}};
         \node [font=\bf,anchor=south west] at (G.north west) {(a) 1 channel};
         \end{tikzpicture}
     \hspace{1.6cm}
         \begin{tikzpicture}
         \node (G) {\includegraphics[width=.49\textwidth]{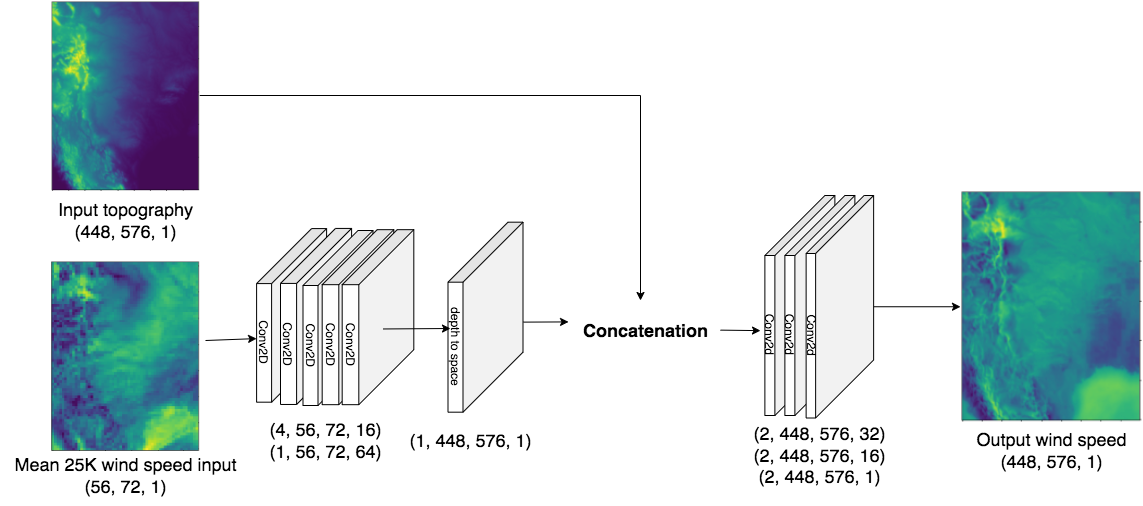}};
         \node [font=\bf,anchor=south west] at (G.north west) {(b) 2 channel};
         \end{tikzpicture}
     \end{subfigure}
     \begin{subfigure}[b]{\textwidth}
         \centering
         \begin{tikzpicture}
         \node (G) {\includegraphics[width=0.48\textwidth]{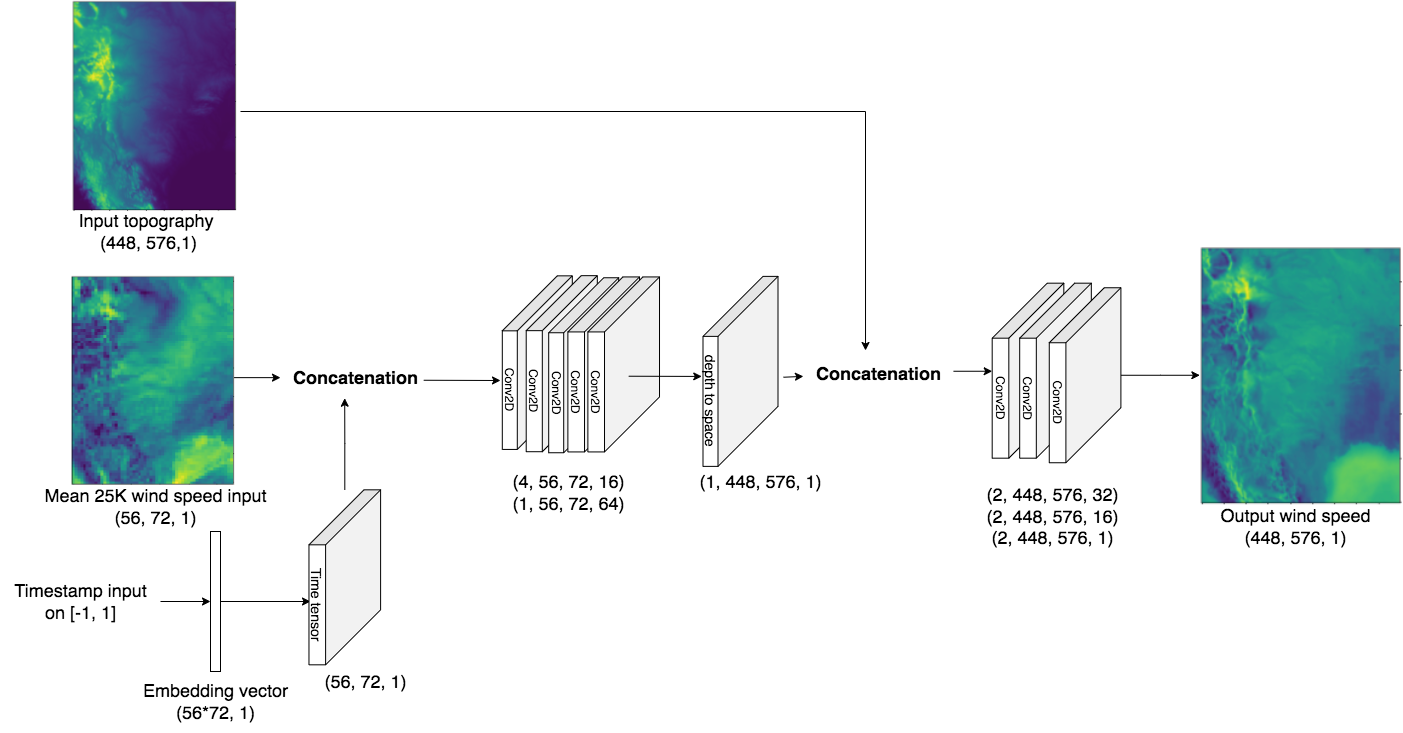}};
         \node [font=\bf,anchor=south west] at (G.north west) {(c) 3 channel};
         \end{tikzpicture}
     \hfill
         \begin{tikzpicture}
         \node (G) {\includegraphics[width=.49\textwidth,height=\textheight,keepaspectratio]{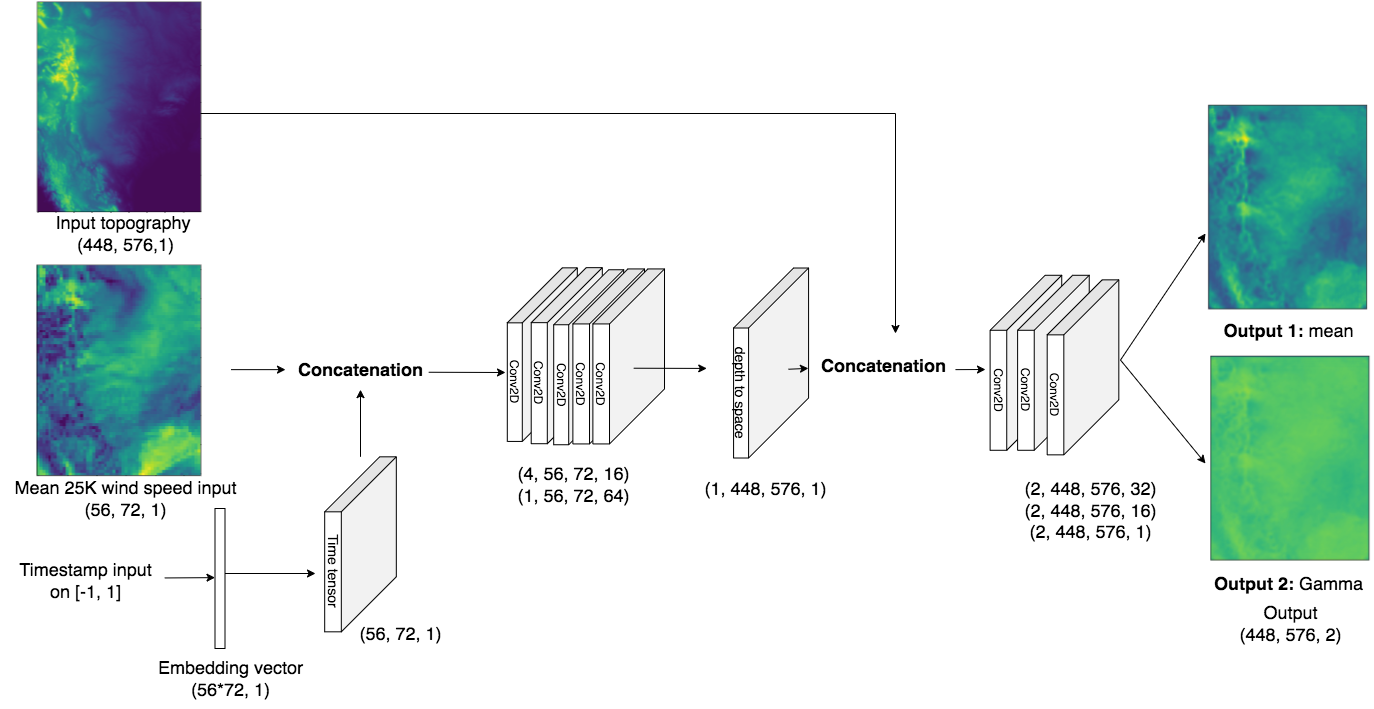}};
         \node [font=\bf,anchor=south west] at (G.north west) {(d) Gumbel};
         \end{tikzpicture}
     \end{subfigure}
     \caption{Architectures of our various SR-CNN configurations. Only 25K model configurations are depicted for space reasons; architectures are similar at higher-resolution gaps but contain more convolution layers in proportion to the downscaling factor.} 
     \label{fig:CNNs}
\end{sidewaysfigure}

This class of models takes fine-scale topography, at 3 km resolution, as a second input (in addition to coarse winds). The topography input informs the model of fine-scale features about the topography that has influential effects on wind speeds.
As in the 1-channel case, the input coarse wind speed is first convolved and downscaled so that it is at the same 3 km-resolution as the topographical data. The downscaled wind speed is then concatenated with topography, creating two channels, and this two-channel tensor is further convolved before the final output layer (see Figure~\ref{fig:CNNs}(b)).

In the 48K and 100K resolution cases, we implement an  alternative variant of the 2-channel setup that involves an initial output at an intermediate resolution (e.g., 48K input has 25K intermediate output) that precedes the final, high-resolution output. 
This intermediate output is equipped with a loss function for  training  the SR-CNN model.
In this configuration, two sets of gradients are computed during model training, because of the two loss functions. The expectation is that targeting convolutional layers earlier in the CNN with one set of gradients will have the effect of fine-tuning model performance across larger resolution gaps.

\subsubsection{3-channel input -- coarse wind speed, topography, time}
This model takes additional temporal information compared with the 2-channel model. 
The temporal input informs the SR-CNN model about diurnal cycles. SCREAM generates a new wind speed field measurement every 15 minutes, so each piece of data is associated with a discrete timestamp within $[0,96)$ corresponding to 15-minute intervals of 24 hours. To impose periodicity on these timestamps, we preprocess the timestamps by passing them through a sine function with a period of 96 serving as a way of encoding the diurnal features of wind speed. The resulting float value is then cast to a tensor the size of the corresponding coarse wind speed input, to which it is concatenated and subsequently passed through the rest of the CNN (see Figure~\ref{fig:CNNs}(c)). We expect this variable to inform the model of differences in intensity and variability of wind speed distributions across the day.

\subsection{Probabilistic loss model}\label{sec:prob_loss}

The preceding SR-CNN models all have in common that they
make deterministic predictions of fine-scale wind speeds. Here, we introduce a different type of model with multiple output layers, namely, 2D maps of probability distribution parameters. The resulting model outputs are parameters of a probabilistic loss function, and these are minimized during training. In particular, we output the location $\mu$ and scale $\sigma$ parameters of a Gumbel distribution. 
The loss function used to train this model is given by the negative logarithm of the Gumbel probability density function (pdf): 
\[L_{gumbel}(\textbf{x}; \mu, \gamma) =  \frac{\textbf{x}-\mu}{\gamma} + e^{-\frac{\textbf{x}-\mu}{\gamma}}\] 
with $\mu$ the mean, $\gamma$ the scale parameter, and \textbf{x}  as our ground-truth sample. 
The outputted parameter $\mu$ has the same interpretation as the previous CNN outputs since it corresponds to a mean/location parameter of the Gumbel pdf. 
The $\gamma$ layer is equipped with the activation $1 + ELU(x)$,
with the exponential linear unit (ELU) activation function:
\[ ELU(x) = \begin{cases} 
      x  & x\geq 0, \\
       e^x - 1 & x\leq 0.\\
   \end{cases}
\]

These parameter maps can then be used to generate samples from a unique pdf at each 3 km point in the region of interest and compute analytical quantiles of the pointwise wind speed distribution. Additionally, this representation enables us to account for the inherent stochasticity and variability of the SWS at each grid point.  
In the same vein, \cite{guillaumin2021stochastic} used a Gaussian probability density function as a loss for a CNN parameterization of ocean momentum forcing; however, Gaussian distributions are known for not capturing extremes, so we  use a Gumbel distribution that is a  particular case of the generalized extreme value distribution. 
We note that the Weibull distribution may have been a natural choice for the loss, since this pdf loss has been widely used to encompass the stochastic variability of wind speed at a given grid-point (see more details in Section \ref{sec:benchmark_model}). However,  to output a wind field from the neural network, one needs to use a pdf loss with a location or mean parameter in the pdf expression, hence limiting the use of a Weibull pdf in this context.

\subsection{Model configuration and training}
\begin{figure}
\centering
\includegraphics[width=1\textwidth]{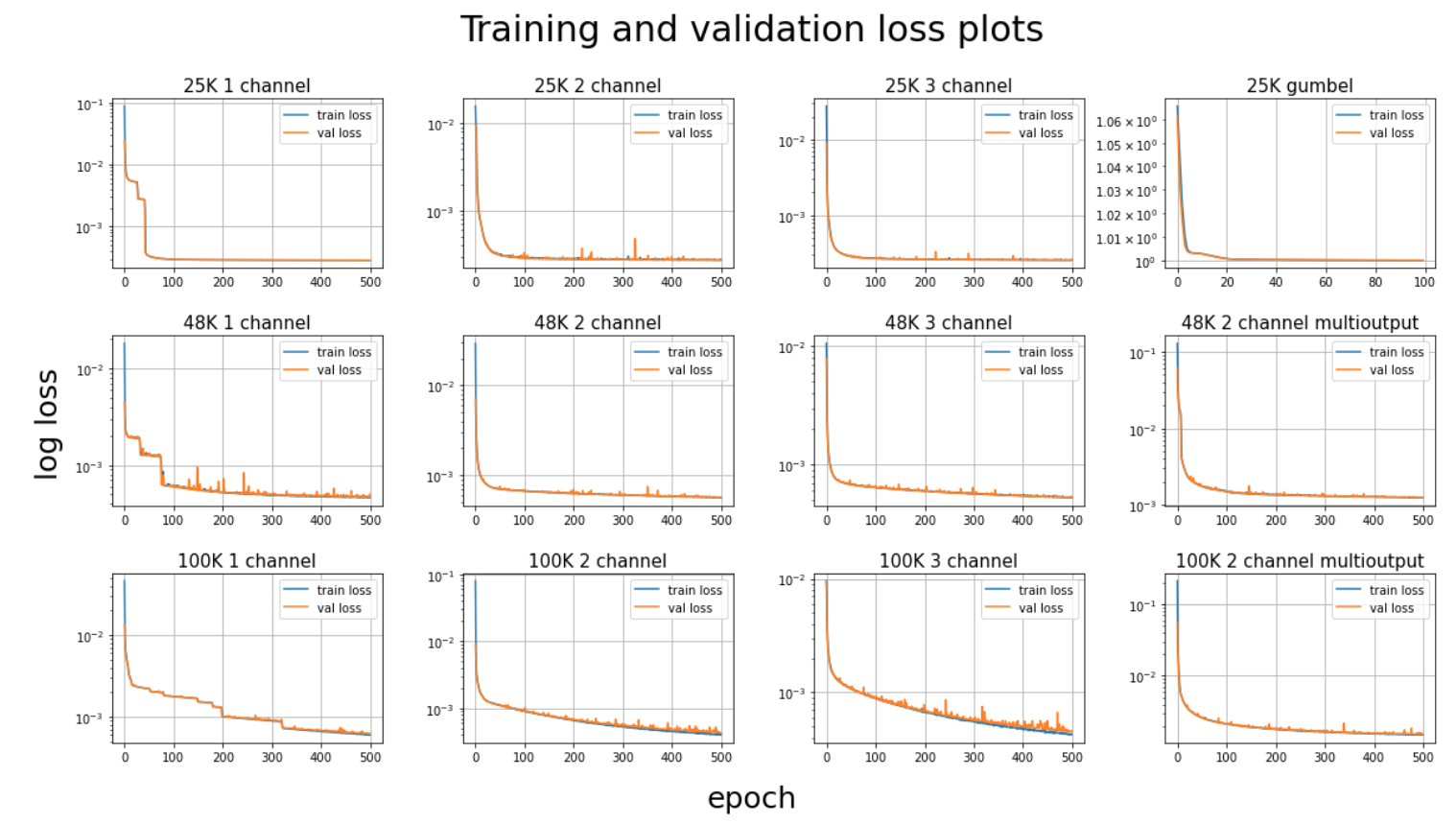}
\caption{Training losses (blue curves) and validation losses (orange curves) in logarithmic scale for SR-CNN models. With the exception of certain spikes in validation loss, both training and validation losses exhibit stable monotonic descent across all epochs, indicating  no signs of overfitting. Validation loss plots are derived from test data that is not used to train models or tune hyperparameters.}
\label{fig:train_val} 
\end{figure}

All hidden network layers across all model configurations were equipped with the ReLU activation function; all models were compiled with the Adam optimizer \cite{KingmaBa15}. Each deterministic model configuration was trained using 80\% of the U.S. data, on 1 GPU for 500 epochs, translating roughly to 15--50 minutes of wall time depending on the downscaling factor. Our Gumbel CNN was trained for only 100 epochs because of faster convergence. All models were tested on a test subset of the U.S. data not used in training,  as well as on all data from the eastern Sahara and eastern Asia regions. Model training and validation loss curves are shown in Figure~\ref{fig:train_val}. We highlight the greater smoothness and stability of our 2-channel model training compared with 1-channel CNNs (see graphs in second column of Figure~\ref{fig:train_val}). 

\section{Assessing Model Performance}\label{sec:assessment}

\subsection{Assessment configurations}
We perform visual and quantitative assessments of the super-resolved wind fields and their restitution of the subgrid-scale ground-truth wind speed. 
Specifically, CNN model predictions, having a resolution of 3 km, are assessed at grid cells of 100 km$\times$100 km in a variety of ways to focus on subgrid-scale variability at the 100 km grid-cell level. 
We initially examine similarity in downscaled predictions to ground-truth wind speeds, in addition to a model's ability to specifically predict upper and lower extremes. Prediction-error percentage and probabilistic metrics at a selection of grid cells of interest are used to assess quantitatively the model efficacy.
In the case of the SR-CNN with probabilistic loss, we use the continuous rank probability score (CRPS) to assess distribution samples at 3 km grid points and the associated ground-truth wind speed.

\subsubsection{Subgrid-scale grid-cell assessment}
For ease of direct comparability, we are interested in how well the models predict subgrid wind speeds at 3 km grouped into 100 km$\times$100 km grid cells, which approximate to the horizontal resolution of the existing global climate models including the standard E3SM. 
 We take 10 random predictions the model performed on test  data (in each of our regions: southeastern U.S., eastern Sahara, eastern Asia) and parse each into approximate 100 km grid-cell subsections. From there, we compare the model predictions to the ground-truth data, paying special attention to grid cells at varying percentiles (minimum 25\%, median 75\%, and maximum) of pointwise prediction error.
Figures \ref{fig:25K grid cells}, \ref{fig:48K grid cells}, and \ref{fig:100K grid cells} show the plots generated from this assessment. Figure \ref{fig:percentile_markers} details an example of where certain grid cells of interest are situated in the United States.

\subsection{Metrics}

\subsubsection{Mean absolute percentage error}
We use as a  metric the mean absolute percentage error (MAPE) and examine various percentiles thereof for intermodel comparison across resolution differences. MAPE is given by \[ \text{MAPE}(y_{true}, y_{pred}) = \frac{1}{n}\sum_{i=1}^{n}\left\vert\frac{y_{true, i}-y_{pred, i}}{y_{true, i}}\right\vert\], where $n$ refers to the number of 3 km points within a 100x100 km grid cell.

\subsubsection{Jensen--Shannon distance}
A key aspect of our model that we want to evaluate is the accuracy and variability of model predictions at 3 km compared with our 3 km ground-truth wind speed data within a 100 km grid cell. To that end, we use the Jensen--Shannon (JS) metric, which measures how alike two probability distributions are. The JS distance is a symmetrized version of Kullback--Leibler (KL) divergence \cite{kldpaper}; and unlike other probability distribution metrics whose ranges are partly unbounded (e.g., Wasserstein metric), the JS distance is real-valued on $[0,1]$, allowing for more intuitive interpretations of results. Values closer to 0 indicate that the two distributions are more alike than not. For two distributions $P$ and $Q$, the JS distance is defined by
\[ \text{JS}(P,Q) = \sqrt{\frac{\text{KL}(P||M) + \text{KL}(Q||M)}{2}}, \] 
where $\text{KL}(\cdot||\cdot)$  refers to the KL divergence between two distributions and $M$ here is the mixture of $P$ and $Q$ distributions with equal contributing probabilities.

\subsubsection{CRPS for Gumbel samples}
In the case of the CNN with probabilistic loss, we use the continuous rank probability score (CRPS) \cite{crpspaper} to assess how well the pointwise Gumbel pdf samples perform relative to deterministic ground-truth data. The CRPS compares each value of a probabilistic sample with a single point according to 
\[CRPS(F, x) = \int_{-\infty}^{+\infty}(F(y) - H(y-x))^2dy,\] 
where $F$ is the cumulative distribution function of the Gumbel sample and $H$ is the Heaviside step function.

\subsection{Benchmark models}\label{sec:benchmark_model}
\subsubsection{Mixture of Weibull distributions for SGS distribution}
We compare subgrid-scale variability histograms and JS values of ground-truth data and model predictions with those computed between ground truth and a  mixture of Weibull distributions estimated numerically by maximum likelihood estimation (MLE) on the data using \cite{reliabilitypackage}. Single Weibull distributions have been shown to accurately capture marginal distributions of surface wind speeds as they accommodate the right skewness of wind speed distributions \cite{Brown84,monahan2006probability, solari2016}. Weibull distributions were also used to represent the subgrid-scale variability of wind speed \cite{zhang2016quantifying}.  In general, however, wind speed distributions often exhibit multimodality that can be due to weather regimes when considering temporal distributions.  
In the present case, subgrid-scale distributions of wind speed (spatial distribution of fine-scale wind speed over coarser grid cell) exhibit some multimodality due to the high resolution of the data capturing various wind-speed behaviors over complex terrains. 
To embed the multimodality due to complex terrain, we extend the single Weibull to a mixture model to better fit to these complex characteristics. We use this MLE fit as an additional baseline to assess how well the CNN models capture wind speed subgrid-scale variability. We note that Weibull mixtures are fitted at the 100 km grid level without accounting for spatial correlation and without providing regional information. 

\subsubsection{Classical interpolation for super-resolution}

We compare our CNN models against a bicubic interpolation baseline based on the implementation of \cite{kurinchi2021} at each of the three resolution differences and each region. 

\section{Results and Discussion}\label{sec:results} 

In this section we assess visually and quantitatively the different proposed models in an out-of-sample fashion. We first evaluate the quality of the super-resolved wind fields over the U.S. region on testing fields (unused during training) when the models are trained over the United States and over the  eastern Sahara and eastern Asia, while the models are trained on U.S. wind fields, too. 
We focus on subgrid-scale variability aspects and, in particular, pay attention to capturing lower and upper parts of the subgrid-scale distributions.

\subsection{U.S. predictions}
\subsubsection{Assessment of super-resolved wind speed errors}

\begin{figure}
\centering
     \begin{subfigure}[b]{\textwidth}
         \centering
         \begin{tikzpicture}
         \node (G) {\includegraphics[width=.31\textwidth]{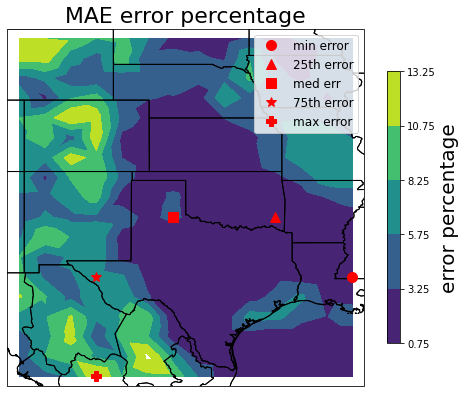}};
         \node [font=\bf, anchor=south west] at (G.north west) {(a) 25K};
         \end{tikzpicture}
         \begin{tikzpicture}
         \node (G) {\includegraphics[width=.31\textwidth]{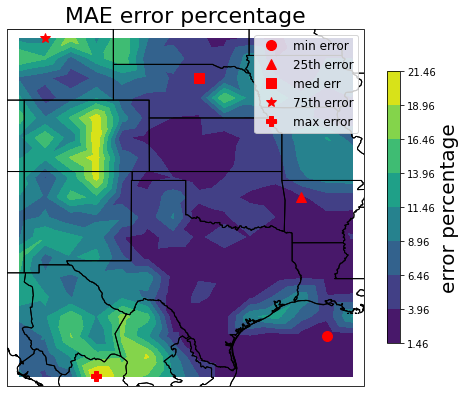}};
         \node [font=\bf,anchor=south west] at (G.north west) {(b) 48K};
         \end{tikzpicture}
         \begin{tikzpicture}
         \node (G) {\includegraphics[width=.31\textwidth]{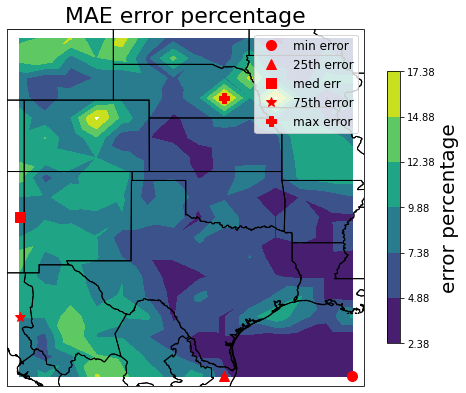}};
         \node [font=\bf, anchor=south west] at (G.north west) {(c) 100K};
         \end{tikzpicture}
     \end{subfigure}
\caption{Maps of mean absolute percentage error  with red markers corresponding to  grid cells with different quantiles of prediction error (minimum, 25th quantile, 50th quantile, 75th quantile, and maximum percentage error) for the same test image. The images correspond to predictions from 2-channel models at 25K (left), 48K (middle), and 100K (right).}
\label{fig:percentile_markers} 
\end{figure}
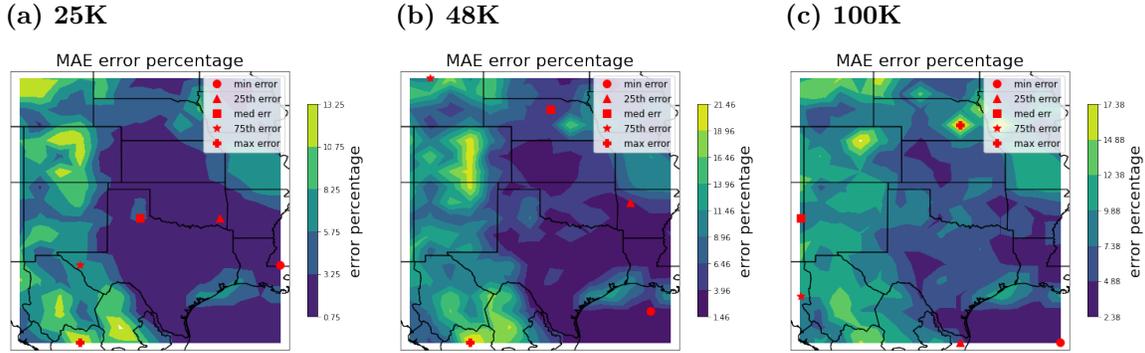

Figure~\ref{fig:percentile_markers} depicts the mean absolute percentage error of super-resolved wind speed highlighting areas of lower and higher quantiles of prediction errors. Areas of lower prediction error are associated with flat terrestrial areas in the region, as well as the Gulf of Mexico. Wind speeds are more homogeneously distributed in these areas, particularly over the ocean \cite{breckling2012}. Higher prediction error can be seen in coastal regions and in the mountains (i.e., parts of the U.S. with more complex terrain).  
The highlighted locations in red with different quantiles of prediction error are used in the following for further evaluations. 
We also note that the errors are larger for greater downscaling factors, where the downscaling factors are 8x (left), 16x (middle), and 32x (right) in Figure~\ref{fig:percentile_markers}, following natural expectations that the greater the downscaling factor, the more challenging the super-resolution.

\begin{figure}
\centering
\includegraphics[width=0.6\textwidth]{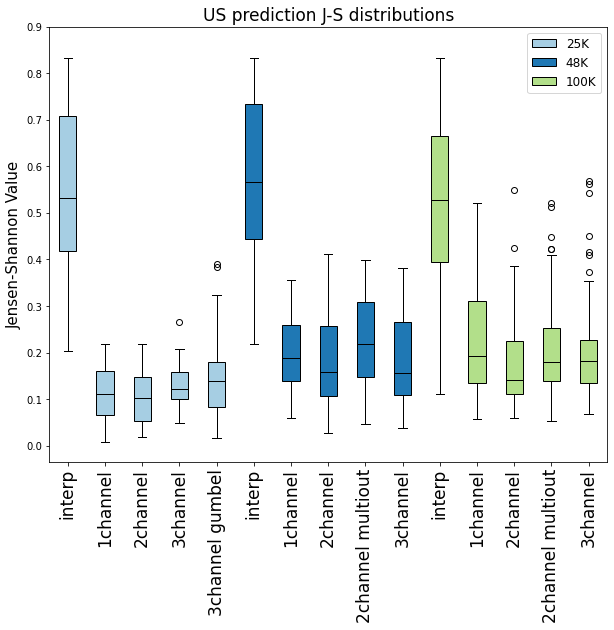}
\caption{Boxplots of Jensen--Shannon value distributions for each model configuration across the 10 test image samples. Jensen--Shannon values are computed between distributions of 3 km super-resolved wind speed and 3 km ground-truth wind speed over 100 km grid cells.  The 25K distributions (yellow) exhibit the lowest median and spread, while 48K (green) and 100K (red) values are centered around higher values with more outliers. Overall, these indicate significant statistical similarity between model predictions and ground-truth values of surface wind speed.}
\label{fig:USJS} 
\end{figure}

\begin{table}
    \centering
    \begin{adjustbox}{width=1\textwidth}
    \begin{tabular}{l|c|ccc|ccc|ccc}
        \textbf{Resolution} & \textbf{Model} & \multicolumn{3}{c|}{{\bf Minimum}} & \multicolumn{3}{c|}{{\bf Median}}  & \multicolumn{3}{c}{{\bf Maximum}}   \\
         & & US & ES & EA & US & ES & EA & US & ES & EA \\\hline
         \textbf{25K} & Interpolation & 26.50\% & 26.97\% & 27.49\% & 34.71\% & 33.62\% & 36.29\% & 81.75\% & 55.43\% & 67.65\% \\
          & 1-channel & 0.36\% & 0.29\% & 1.07\% & 4.12\% & 2.76\% & 9.81\% & 51.89\% & 32.23\%& 37.71\% \\
          & 2-channel & 0.29\% & 0.35\% & 1.09\% & 4.09\% & 2.73\% & 9.62\% & 56.22\% & 31.73\% & 33.50\% \\
          & 3-channel & 0.92\% & 1.17\% & 1.44\% & 4.18\% & 3.58\% & 9.89\% & 57.39\% & 30.97\% & 35.19\% \\
          & Gumbel (mean)  & 0.64\% & & & 4.79\% & & & 78.41\% & & \\\hline
         \textbf{48K} & Interpolation & 15.57\% & 18.55\% &20.89\% & 38.03\% & 35.33\% & 41.50\% & 117.38\% & 71.76\% & 104.23\%\\
          & 1-channel & 0.98\% & 0.90\%&  2.58\% & 6.89\% & 6.13\% & 17.84\% & 84.13\% & 48.40\%& 73.17\% \\
          & 2-channel & 0.93\% & 0.72\% & 2.44\% & 6.99\% & 5.75\% & 16.92\% &  97.20\% & 44.95\% & 81.99\% \\
          & 2-channel multiout & 1.20\% & 1.13\% & 2.57\% & 7.00\% & 5.72\% & 17.14\% & 89.60\%  & 41.87\% & 68.78\% \\
          & 3-channel & 0.83\% & 1.44\% & 5.41\% & 6.81\% & 10.04\% & 24.60\% & 85.51\% & 56.95\% & 72.4\% \\\hline
          \textbf{100K}& Interpolation & 4.29\% & 3.52\% & 8.52\% & 34.33\% & 26.06\% & 39.19\% & 135.72\% & 100.95\% & 141.56\%\\
           & 1-channel & 1.57\% & 2.25\% & 3.38\%& 8.57\%  & 12.12\% & 26.41\% & 113.01\% & 65.00\% & 106.98\%\\
          & 2-channel & 1.65\% &2.28\% & 7.82\% & 7.48\% & 13.81\% & 30.90\% & 79.59\% & 62.00\% & 122.07\% \\
          & 2-channel multiout & 1.66\% & 2.69\% & 6.36\% & 7.42\% & 14.98\% & 38.07\% & 80.68\% & 68.43\% & 177.73\% \\
          & 3-channel & 1.91\% & 2.98\% & 12.34\% & 7.69\% & 18.95\% & 47.02\% & 86.41\% & 97.12\% & 250.92\% \\
    \end{tabular}
    \end{adjustbox}
    \caption{Minimum, median, and maximum MAPE of model prediction error percentage within 100 km grid cells in each region of interest (US, eastern Sahara (ES), and eastern Asia (EA)), compared against interpolation baseline. The figure shows  a general trend in CNN error increasing within each percentile as resolution increases, the exception being the behavior of maximum error in the U.S. region.}
    \label{tab:mean_grid_cell_pred_error}
\end{table}

Table \ref{tab:mean_grid_cell_pred_error} details the mean absolute error percentage for model predictions at the 100 km grid cells of interest (Figure \ref{fig:percentile_markers}). 
Median percentage errors remain overall low over all resolution gaps and regions, indicating realistic super-resolved wind fields. 
Minimum errors are also low except for eastern Asia at higher-resolution gaps. 
We conjecture this exception is due to the complex terrain in eastern Asia and its daily cycles that are more pronounced  than the learned ones over the U.S. data. 
Overall, except in eastern Asia, minimum errors are extremely low, indicating very accurate super-resolved wind speed. 
As in the comparison of JS value distributions, all CNN model configurations outperform the interpolation baselines at all resolutions and error percentiles over the U.S. region. Looking only at the CNN models, we can see a trend of error percentage increasing slightly at higher downscaling factors, particularly in grid cells at minimum and median error percentage quantiles. 
Similar error percentages are seen in the maximum error grid cells between the 48K and 100K models. Looking at the slightly lower median error percentages at 48K reveals that the distribution of these error percentages show higher spread compared with 100K errors. Thus, while similar in value, the 48K models' maximum error values make up less of the overall distribution than do those of the 100K models. 
Overall, prediction errors have a positively skewed distribution with longer right tails as shown by the significantly larger difference between maximum and median errors compared with the difference between minimum and median errors. 
We note that prediction errors  for  eastern Sahara are most of the  time the smallest, highlighting very good prediction accuracy, although this region is not used for training. 
We believe this is attributed to the more homogeneous terrain of that region compared with that of the other regions.

\subsubsection{JS values for subgrid-scale distributions}
Figure~\ref{fig:USJS} presents a high-level overview of the JS values of CNN model predictions compared with ground-truth U.S. data within 100 km grid cells. Each boxplot in the figure describes the distribution of 50 JS values, taken from 5 different grid cells (see Figure \ref{fig:percentile_markers}) across the 10 test samples.
Comparing the distribution of JS values of the CNN models with those of our interpolation baseline, we see that the bulk of the distributions of the former is significantly lower than those of the latter. Although the lower extremes of the interpolation JS values overlap with their CNN counterparts, this metric clearly shows that the CNNs capture wind speed variability more accurately than does the baseline interpolation.

We see a general trend of JS values exhibiting lower medians and overall lower spread at lower-resolution gaps (from left to right in Figure~\ref{fig:USJS}); this is especially noticeable between the 25K (yellow color) and 48K (green color) plots. While the Gumbel model predictions (labeled ``3channel gumbel'') exhibit the largest distribution spread at 25K, the interquartile range is consistent with its counterparts. 
The larger spread of the Gumbel model JS values  highlights that the model performances are more variable from one testing sample to another. 
Below, we further discuss Gumbel model performance. 
We see higher maximum JS values in the 100K model case compared with 48K, as well as outlier values that are not present for the 48K models. This higher spread in JS values is consistent with the idea that downscaling models perform better at smaller resolution gaps.

In Figure~\ref{fig:USJS}, 2-channel models (i.e., coarse wind and fine terrain inputs) exhibit the lowest median JS value within each resolution difference.              
This highlights that more complex neural network models that are informed by richer data are capable of  reducing prediction errors. 
The models with 3 channels (i.e., with added timestamp), however, are comparatively slightly less accurate in their predictions as measured by JS.
During training, these models did not exhibit signs of overfitting (see Figure~\ref{fig:train_val}).
Therefore, we conclude that simply adding additional data may deteriorate predictions if not chosen appropriately and that timestamps may not reflect the physicality of wind speeds.
We discuss potential solutions in the conclusion and perspective section.

\subsubsection{Visual assessment of subgrid-scale variability of super-resolved wind speed}

\begin{figure}
     \centering
     \begin{subfigure}[b]{0.49\textwidth}
         \centering
         \includegraphics[width=\textwidth]{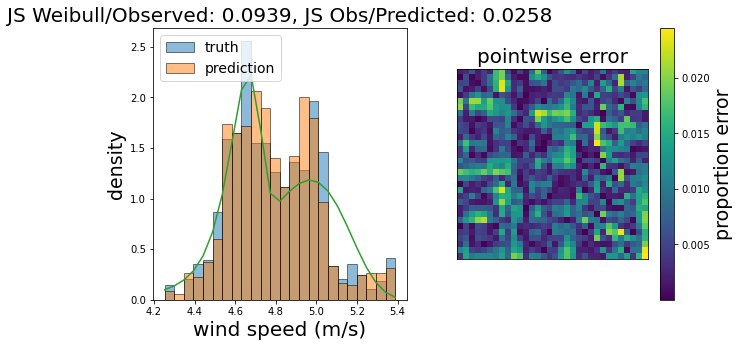}
     \end{subfigure}
     \begin{subfigure}[b]{0.49\textwidth}
         \centering
         \includegraphics[width=\textwidth]{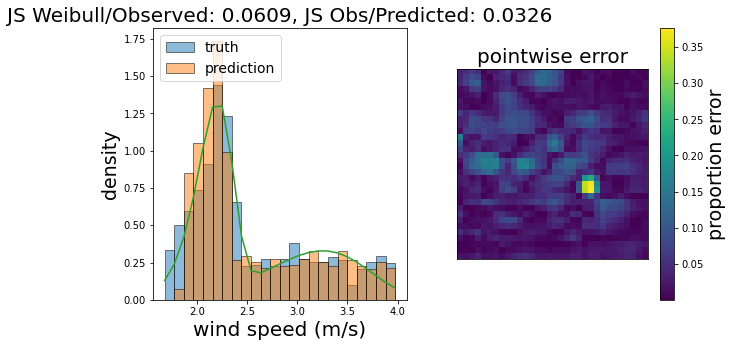}
     \end{subfigure}
     \begin{subfigure}[b]{0.49\textwidth}
         \centering
         \includegraphics[width=\textwidth]{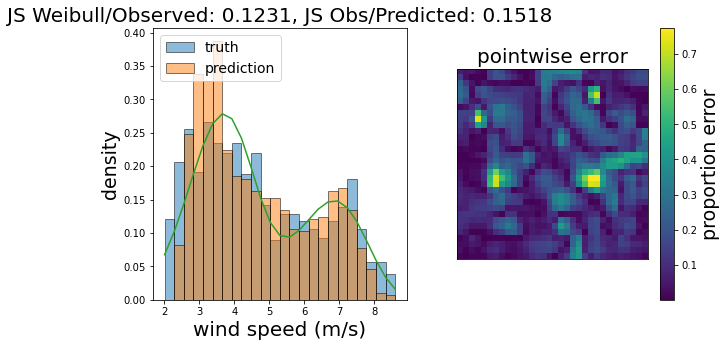}
     \end{subfigure}
        \caption{2-channel 25K CNN prediction comparison to ground-truth data and Weibull MLE fit on SGS distributions (green line) at grid-cell level. Values above the left-hand plot correspond to the JS values between (1) MLE-fitted Weibull mixture and ground-truth data and (2) the model prediction and ground-truth data. The left-hand side of each plot depicts histograms of predictions overlain with ground truth over a grid cell and Weibull mixture pdf; the right side depicts pointwise prediction error proportion within the grid cell.  From  top  left in clockwise order: super-resolved wind  speed in grid cells with minimum error, median, and maximum prediction error.}
        \label{fig:25K grid cells}
\end{figure}

\begin{figure}
     \centering
     \begin{subfigure}[b]{0.49\textwidth}
         \centering
         \includegraphics[width=\textwidth]{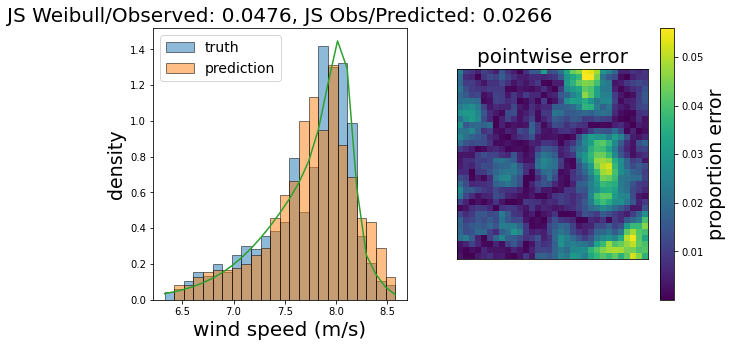}
     \end{subfigure}
     \begin{subfigure}[b]{0.49\textwidth}
         \centering
         \includegraphics[width=\textwidth]{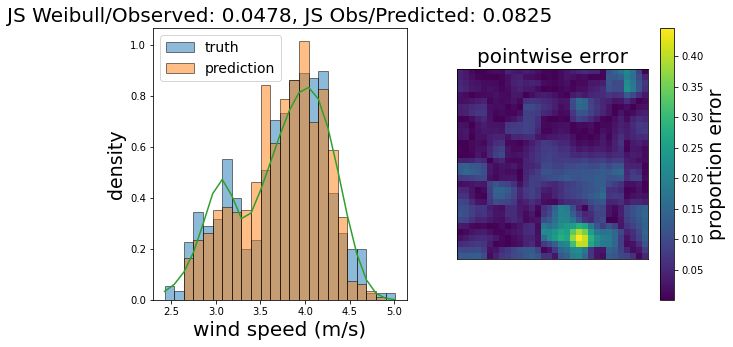}
     \end{subfigure}
     \begin{subfigure}[b]{0.49\textwidth}
         \centering
         \includegraphics[width=\textwidth]{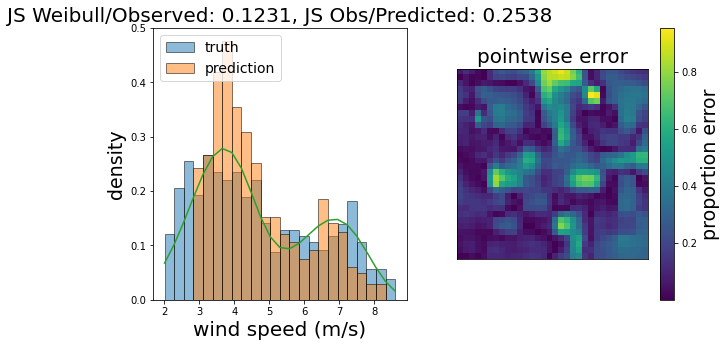}
     \end{subfigure}
        \caption{Same content as Figure \ref{fig:25K grid cells} for 2-channel 48K CNN prediction comparison with ground-truth data and Weibull MLE fits (in green) at grid-cell level.}
        \label{fig:48K grid cells}
\end{figure}

\begin{figure}
     \centering
     \begin{subfigure}[b]{0.49\textwidth}
         \centering
         \includegraphics[width=\textwidth]{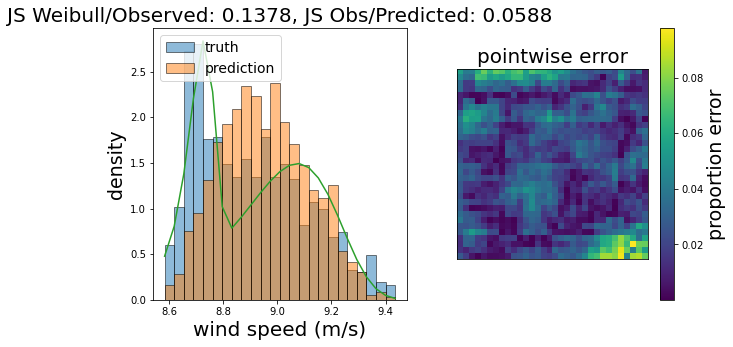}
     \end{subfigure}
     \begin{subfigure}[b]{0.49\textwidth}
         \centering
         \includegraphics[width=\textwidth]{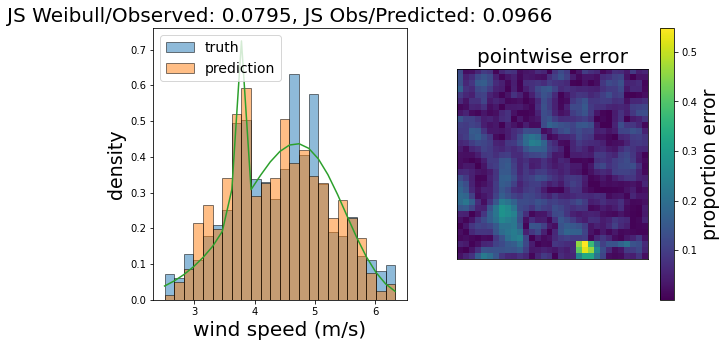}
     \end{subfigure}
     \begin{subfigure}[b]{0.49\textwidth}
         \centering
         \includegraphics[width=\textwidth]{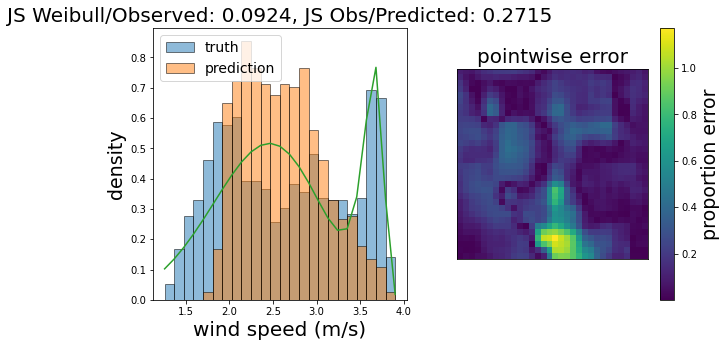}
     \end{subfigure}
        \caption{Same content as Figure \ref{fig:25K grid cells} for 2-channel 100K CNN prediction comparison with ground-truth data and Weibull MLE fits (in green) at grid-cell level.}
        \label{fig:100K grid cells}
\end{figure}

Figures \ref{fig:25K grid cells}, \ref{fig:48K grid cells}, and \ref{fig:100K grid cells} further elaborate on the 2-channel models' ability to capture 3 km wind speed variability at the 100 km grid-cell level. They respectively show results of 3 km super-resolved wind speed from 25 km, 48 km, and 100 km coarse inputs.

Figure \ref{fig:25K grid cells} shows a trend of very low JS values in grid cells at lower percentiles of prediction error. The JS values increase at higher percentiles, as seen in the maximum error plot in the figure, although still comparable to the Weibull mixtures (this is corroborated by Figures \ref{fig:48K grid cells} and \ref{fig:100K grid cells}). Examining the histograms in these figures, we see that  the 2-channel CNNs display a strong ability to predict the bimodality, positive skewness, and tails seen in the true wind speed distributions. In many cases, model predictions capture these characteristics more closely than do the Weibull mixture pdfs. An important note here is that the CNN is making a prediction based on unseen coarsened
wind speed and other predictors, whereas Weibull mixtures are fit directly to the ground-truth test data. Weibull mixtures were also given no additional information about wind speeds in surrounding grid cells and thus are highly local fits. The results demonstrate that our CNNs frequently recover fine-scale wind speeds better than localized Weibull mixture models do. Additionally, even when predictions are slightly worse compared with these mixtures, the CNNs' downscaling process is more general with spatially correlated outputs and is less computationally expensive than performing MLE on 3 km data at specific grid cells. 
Predictions in grid cells of higher error can still capture the major features of true wind distributions well. However, they reflect this error in the form of over/underestimation of modes or shifts in the overall center of the predicted distribution. This is most apparent in the 100K 2-channel CNN.

\subsubsection{US predictions with Gumbel-loss model}\label{sec:gumbel}

\begin{figure}
\centering
\includegraphics[scale=.7]{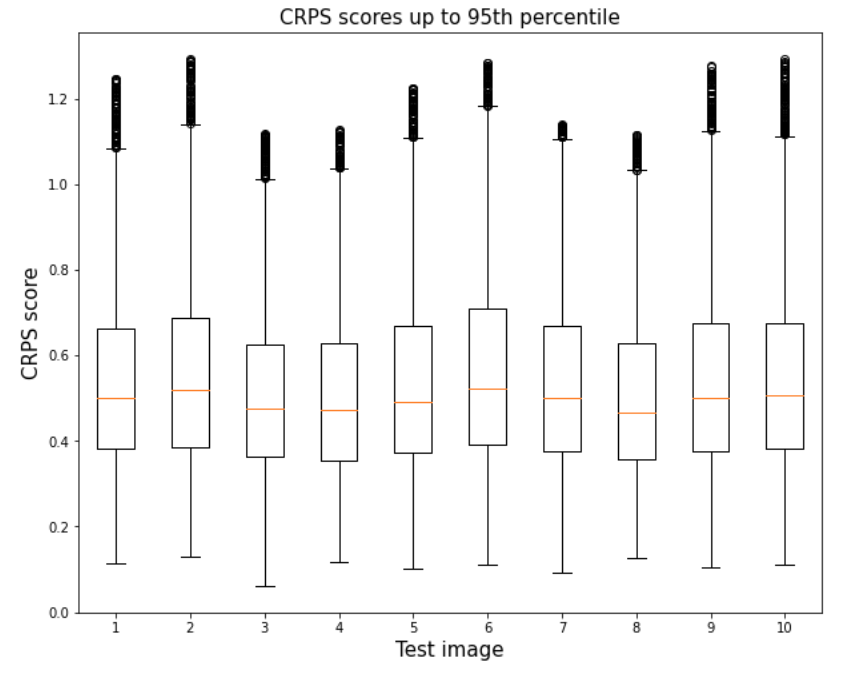} 
\caption{CRPS scores for samples generated by downscaled Gumbel pdf parameters. Scores were computed at 3 km gridpoints across the entire U.S. region, across our 10 test images.}
\label{fig:crps_boxplots}
\end{figure}

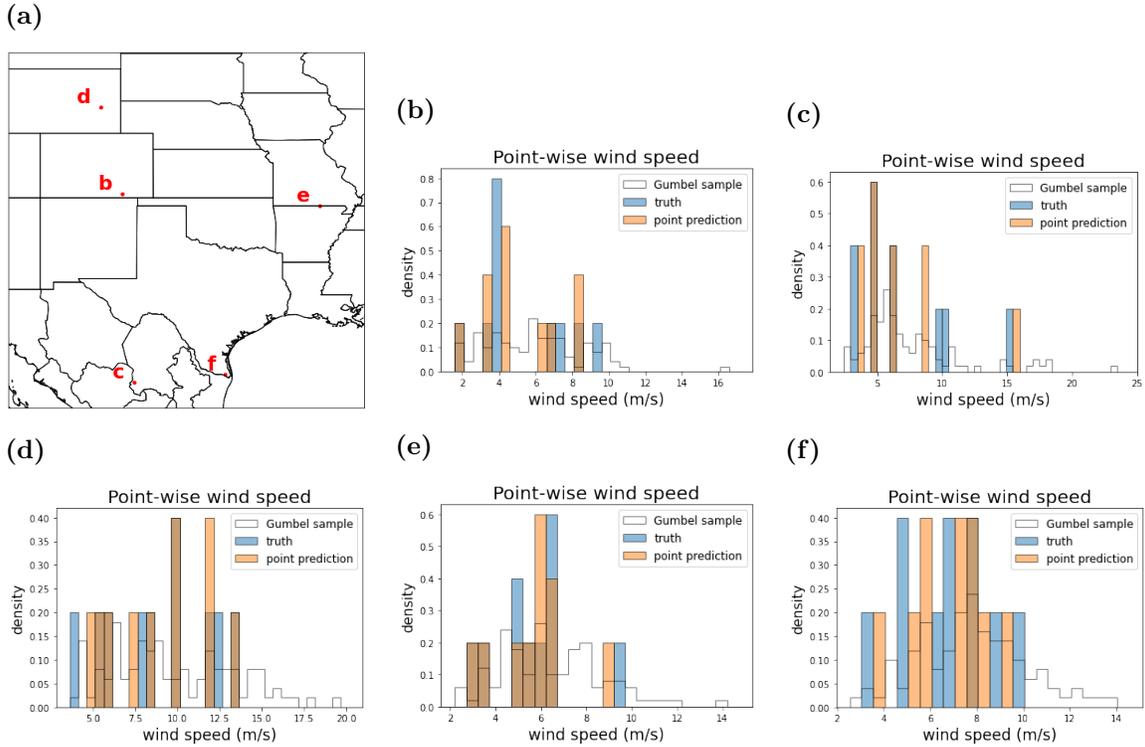
\begin{figure}
    \centering
    \begin{subfigure}[b]{\textwidth}
    \centering 
    \begin{tikzpicture}
    \node (G) {\includegraphics[width=.31\textwidth]{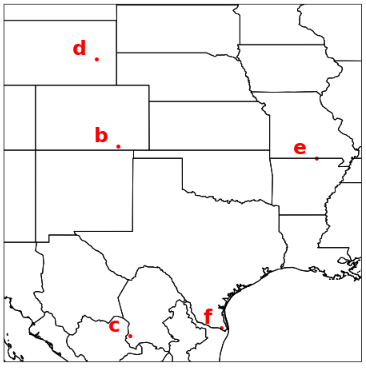}};
    \node [font=\bf,anchor=south west] at (G.north west) {(a)};
    \end{tikzpicture}        
    \begin{tikzpicture}
    \node (G) {\includegraphics[width=.31\textwidth]{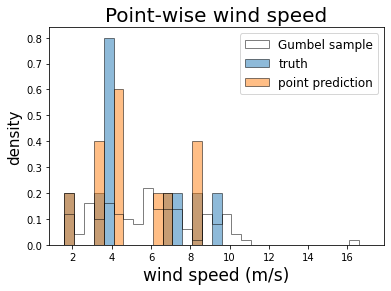}};
    \node [font=\bf,anchor=south west] at (G.north west) {(b)};
    \end{tikzpicture}        
    \begin{tikzpicture}
    \node (G) {\includegraphics[width=.31\textwidth]{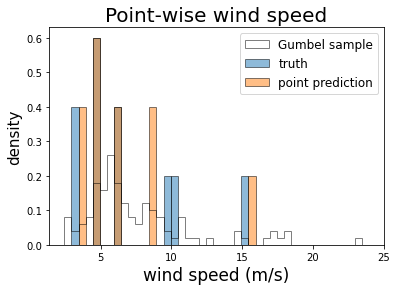}};
    \node [font=\bf,anchor=south west] at (G.north west) {(c)};
    \end{tikzpicture}        
    \begin{tikzpicture}
    \node (G) {\includegraphics[width=.31\textwidth]{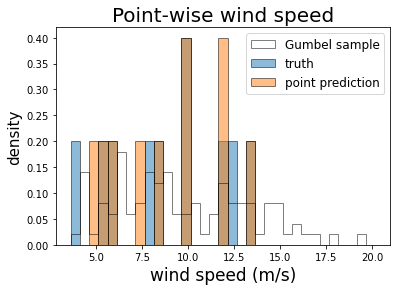}};
    \node [font=\bf,anchor=south west] at (G.north west) {(d)};
    \end{tikzpicture}        
    \begin{tikzpicture}
    \node (G) {\includegraphics[width=.31\textwidth]{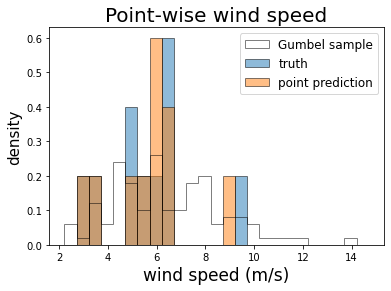}};
    \node [font=\bf,anchor=south west] at (G.north west) {(e)};
    \end{tikzpicture}        
    \begin{tikzpicture}
    \node (G) {\includegraphics[width=.31\textwidth]{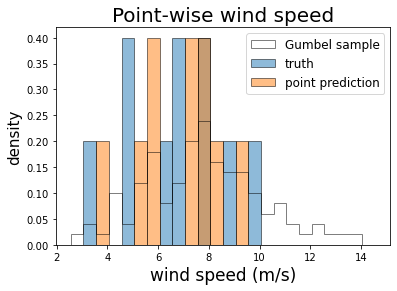}};
    \node [font=\bf,anchor=south west] at (G.north west) {(f)};
    \end{tikzpicture}        
    \end{subfigure}
    \caption{ \textbf{(a)} Five  randomly selected locations of wind speeds described by histograms in \textbf{(b)}--\textbf{(f)}. 
    Histograms show true wind, mean prediction, and aggregated Gumbel distribution samples from the Gumbel CNN over our 10-image test sample.}
    \label{fig:crps}
\end{figure}

The JS values of the Gumbel model predictions in Figure \ref{fig:USJS} (a) were generated  based only on the mean output layer of the model and do not account for the stochasticity provided by the full 3 km pointwise Gumbel pdfs. We examine these further by taking a random sample of 3,000 test points per image in our set of 10, following which we compute the CRPS between each pdf sample at each test point; see the plot of CRPS distributions per sample images  in Figure \ref{fig:crps_boxplots}. 
We find that the majority of the CRPS values up to the 95th quantile fall within the [0,1.2] range, and are similarly distributed across all 10 image samples. A few CRPS scores sit at higher values. Geographically, these points are concentrated either over areas of more complex terrain (and in turn wind speed distribution) or at the boundary of the region. The former speaks to the model's capacity to downscale complex information, while the latter is an issue with CNNs' ability to process information at the edges of images. 
Figure \ref{fig:crps} (b)--(g) detail model performance at five randomly selected 3 km points in the United States, wherein we display Gumbel CNN mean predictions and associated Gumbel samples against true wind data across our 10-image sample. We see that the Gumbel mean values are distributed similarly to the true data, and samples from Gumbel pdfs at these locations accurately capture the spread of  true wind speeds.

\subsection{Eastern Sahara and eastern Asia predictions}

\begin{figure}
\centering
\begin{subfigure}[b]{0.45\textwidth}
\centering
\includegraphics[width=\textwidth]{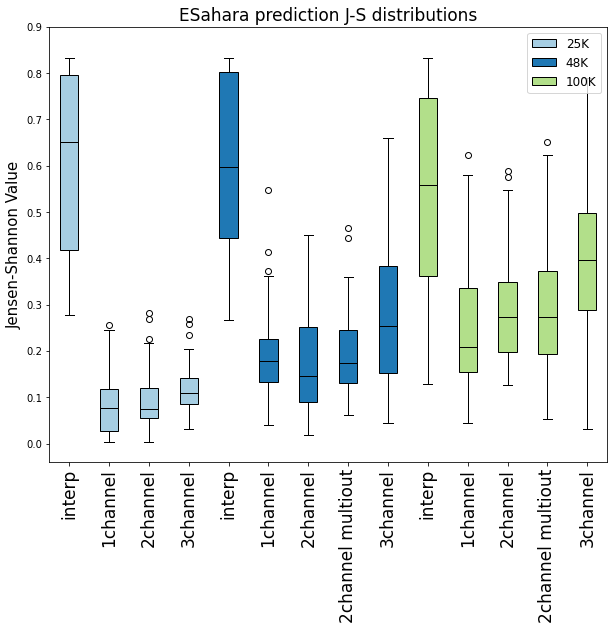}
\caption{JS values over eastern Sahara}
\end{subfigure}
\hfill
\begin{subfigure}[b]{0.45\textwidth}
\centering
\includegraphics[width=\textwidth]{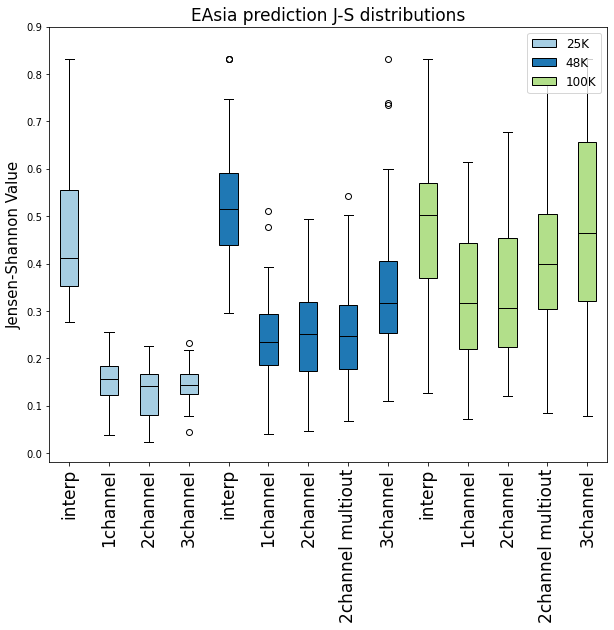}
\caption{JS values over eastern Asia}
\end{subfigure}
\caption{Boxplots of Jensen--Shannon value distributions for each model configuration across the 10-image eastern Sahara and eastern Asia sample.}
\label{fig:ES_EA_JS} 
\end{figure}

\subsubsection{JS values for subgrid-scale distributions }
Figure \ref{fig:ES_EA_JS} highlights the same JS statistics as Figure \ref{fig:USJS} but for CNN predictions over the eastern Sahara and eastern Asia regions. Unlike U.S. predictions, which are performed on a subset of images from the original training region (not training set), here the CNNs make predictions that, while still conditioned on coarsened wind speed, are over regions completely unseen by the models. Overall, JS values remain competitive  with the JS  values of Figure \ref{fig:USJS} given the out-of-sample setting. 
As in Figure \ref{fig:USJS}, we see better JS distributions from CNN predictions compared with the interpolation baseline, although  less pronounced in the eastern Asia region. 
In both plots, we see a more pronounced trend of JS distributions increasing in both median and spread as the  downscaling resolution gap increases. While the performances of the 25K models are similar to those over the United States, the same cannot be said for the 48K and 100K models,  particularly  for the timestamp models. A possible explanation lies in our temporal embedding being based on a single time stamp, which may not be granular enough to account for the variability of diurnal wind cycles within a region. 
In addition, as observed on Figure \ref{fig:hourly_wind}, diurnal patterns are more pronounced in the eastern Asia region than in the U.S. region that is used to train the model, hence a potential lack of adaptability in this region. 
Based also on Figure \ref{fig:hourly_wind} \textbf{(f)}, the higher spread in the  time-averaged wind speed distribution over eastern Asia compared with other regions could also explain poorer performance at 100K. However, this must be considered in conjunction with the 100K eastern Sahara predictions, which display higher JS values relative to other models at smaller downscaling gaps, yet exhibit much lower spatial variability.

\subsubsection{Comparing super-resolved wind speed errors}\label{sec: OOS_error}

Looking again at Table \ref{tab:mean_grid_cell_pred_error}, we see a pattern in the prediction error consistent with Figure \ref{fig:ES_EA_JS}'s JS values. We highlight that the majority of CNN models' out-of-sample error percentages are far lower than those of the interpolation baseline. The main exceptions appear to be in the 100K models, with similar maximum error values over the eastern Asia region and similar minimum error over the eastern Sahara region between CNN and the interpolation model. 
The models' MAPE over the eastern Sahara region consistently increases across larger resolution gaps, which is particularly notable when looking at the median and max error values. In terms of error, the 48K models appear to generalize better to the unseen region than do those in the 100K case, while the 25K models still perform best. There also appears to be lower maximum prediction error across all resolutions over the eastern Sahara when compared with the United States, while median error values do  worse than the United States at the 3-channel 48K model, as well as all 100K models. This likely means there were fewer anomalous predictions in this region than in the United States, possibly due to more homogeneous topography and wind speeds in the Sahara. 
eastern Asia predictions exhibit some of the highest error percentages at all percentiles and resolution differences except the 25K maximum values. Maximum percentage values are comparable to the other regions' values at 48K resolution and are categorically higher than other regions at 100K resolution. When we compared these with the eastern Sahara percentage values, we posit that the complex wind speed distribution over eastern Asia is difficult to downscale for our models. Were this an issue of overfitting, we would expect to see similar poor performance in the eastern Sahara predictions, in both JS value and error percentage.

\subsection{Extreme value assessment}
\subsubsection{Deterministic CNN performance}
\begin{table}
    \centering
    \begin{adjustbox}{width=1\textwidth}
    \begin{tabular}{l|c|cc|cc|cc}
        \textbf{Resolution} & \textbf{Model} &  \multicolumn{2}{c|}{{\bf US}} &   \multicolumn{2}{c|}{{\bf \textbf{Eastern Sahara}}} &             \multicolumn{2}{c}{{\bf \textbf{Eastern Asia}}} \\ 
        \textbf{} & & $\leq$5th & $\geq$95th & $\leq$5th & $\geq$95th & $\leq$5th & $\geq$95th \\\hline
        \textbf{25K} & Interpolation & 78.96\% & 0.0\% & 80.71\% & 0.0\% & 52.88\% & 0.05\% \\
         & 1-channel & 3.67\% & 3.28\% & 4.19\% & 4.10\% & 2.99\% & 2.98\% \\
         & 2-channel & 2.49\% & 4.70\% & 3.44\% & 4.02\% & 3.06\% & 3.54\% \\
         & 3-channel & 3.84\% & 7.00\% & 2.55\% & 7.83\% & 3.00\% & 4.38\%\\\hline
         \textbf{48K} & Interpolation & 83.32\% & 0.0\% & 94.78\% & 0.0\% & 76.14\% & 0.15\% \\
          & 1-channel & 1.57\% & 1.39\% & 4.35\% & 7.28\% & 2.73\% & 1.03\%\\
         & 2-channel & 2.60\% & 3.30\% & 3.15\% & 3.82\% & 0.39\% & 1.55\%\\
         & 2-channel multiout & 3.11\% & 1.33\% & 3.32\% & 1.60\% & 1.02\% & 3.32\%\\
         & 3-channel & 3.25\% & 2.52\% & 17.24\% & 1.95\% & 3.52\% & 11.41\%\\\hline
         \textbf{100K} & Interpolation & 73.22\% & 2.68\% & 71.11\% & 7.77\% & 60.82\% & 4.79\%\\
          & 1-channel & 1.31\% & 3.77\% & 11.87\% & 5.82\% & 7.46\% & 7.87\%\\
          & 2-channel & 4.39\% &  5.49\% & 20.00\% & 8.18\% & 0.71\% & 20.19\%\\
          & 2-channel multiout & 2.66\% & 3.87\% & 15.99\% & 4.91\% & 1.52\% & 13.50\%\\
          & 3-channel & 3.52\% & 1.55\% & 18.10\% & 0.43\% & 3.50\% & 26.28\%\\
    \end{tabular}
    \end{adjustbox}
    \caption{Percentage of model predictions whose values are above the 95th quantile and below the 5th quantile value of true wind speed. The number of downscaled points satisfying this threshold was aggregated from the grid cells of median prediction error across our 10-image test sample. A ``perfect" model would achieve 5\% at both thresholds; here, many models across all resolutions come very close to that percentage, most notably the 2-channel models.}
    \label{tab:extreme_pct}
\end{table}

Table \ref{tab:extreme_pct} provides insights into the performance of the CNN models regarding extreme wind speed prediction. We show that CNN models' extreme performance is favorable to that of our interpolation baseline. Interpolation serially overestimates lower extreme values, almost completely missing upper extreme wind speeds. 
Figures \ref{fig:25K grid cells}, \ref{fig:48K grid cells}, and \ref{fig:100K grid cells} show that in areas of higher error, our models slightly underestimate or overestimate extreme values. Using the 5th and 95th quantiles of ground-truth wind speed in each grid cell of median prediction error of our 10 image testing sample, we see in this table  how close model predictions at either extreme come to this threshold. Many models' extreme value outputs sit very closely to the 5\% mark, most frequently in the case of the 25K models over all regions, but also the 2-channel 48K model over the U.S. and eastern Sahara regions and the 2-channel 100K model over the U.S. region. 
Over the U.S.  region, the 2-channel models at all resolutions appear to capture the upper extremes the best, although  their lower extreme predictions are outperformed by other models at 25K and 48K resolutions. 
When models at 25K and 48K overpredict, it is always upper extreme values. We see most marked overprediction in the eastern Sahara and eastern Asia regions at both extremes in the 100K model case. Specifically, there appears to be overprediction of lower extremes over eastern Sahara and of upper extremes over eastern Asia. 
Overall, we find no demonstrated pattern of mutual exclusivity of extreme prediction, in that we do not consistently see high overprediction in upper extreme wind speed coupled with underprediction in the lower extreme, or vice versa.

\begin{figure}
    \centering
    \hspace*{\fill}
    \begin{subfigure}[b]{0.4\textwidth}
        \centering
        \includegraphics[width=\textwidth]{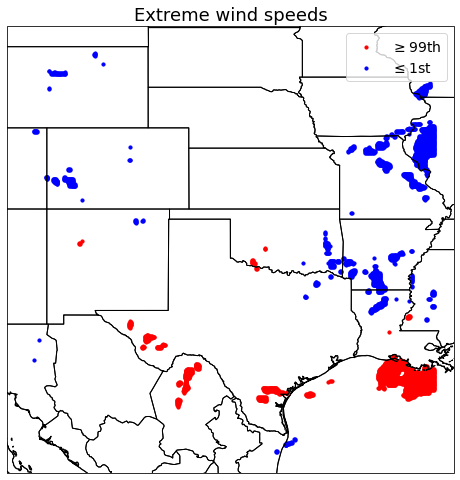}
        \caption{}
    \end{subfigure} \hfill
    \begin{subfigure}[b]{0.43\textwidth}
        \centering
        \includegraphics[width=\textwidth,height=6.2cm]{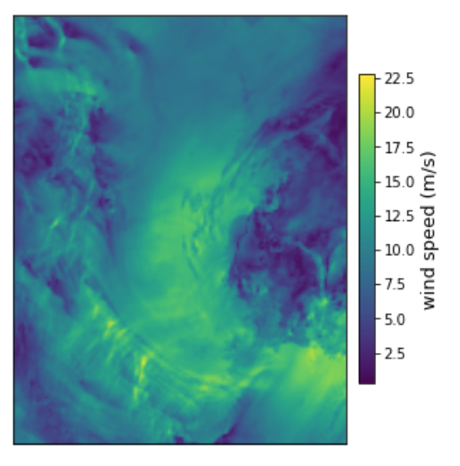}
        \caption{}
    \end{subfigure} \\
    \begin{subfigure}[b]{0.45\textwidth}
        \centering
        \includegraphics[width=\textwidth]{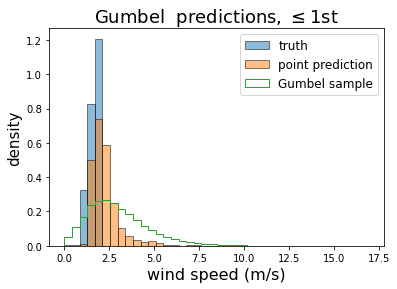}
        \caption{}
    \end{subfigure}
    \begin{subfigure}[b]{0.45\textwidth}
        \centering
        \includegraphics[width=\textwidth]{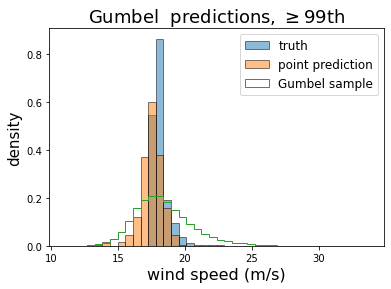}
        \caption{}
    \end{subfigure}
    \caption{\textbf{(a)} Location of true wind speeds at either extreme (blue: points below 1st quantile, red: points above 99th quantile) for a test image with highest spatial mean in our 10-image U.S. sample. \textbf{(b)} Overall wind speed distribution over entire U.S. at same time-snapshot as panel \textbf{(a)}. 
\textbf{(c)--(d)} Gumbel CNN mean predictions (orange) and combined samples generated (green) by unique pdfs at geolocations of true extremes (blue).}
    \label{fig:gumb_extremes}
\end{figure}

\subsubsection{Gumbel-loss CNN performance}

We examine the Gumbel CNN's performance at low and high wind extremes of the test image with highest spatial mean and spatial variability in Figure \ref{fig:gumb_extremes} as a proxy for a case with the most extremes in the selected fields. The mean layer output is compared directly with the deterministic ground-truth wind speeds at either extreme as shown in (c) and (d) of the figure. The model-predicted mean captures the approximate location of true wind as shown but exhibits slight overestimation (underestimation) of lower (upper) extremes. These results must be considered in conjunction with samples taken from Gumbel pdfs at each 3 km location since this model is not deterministic. Distributions of these samples are shown in green in Figure \ref{fig:gumb_extremes}. The true wind speeds fall well within the bulk of the aggregate samples, which exhibit positive skewness at both extremes. Such skewness is seen more in the upper extreme ground-truth wind speeds; the lower extreme samples follow the predicted means of the CNN. The results indicate that alternate modes of model training are required to produce improved scale parameter predictions. Overall, this Gumbel CNN shows promise as a sampling mechanism to generate realistic fine-scale wind speed values in a downscaling context.

\section{Conclusion and Perspectives}\label{sec:conclusion}

\subsection{SR-CNN performances}
In this study we implemented a variety of SR-CNN models for downscaling wind speed fields and examined their performance at different resolution gaps, with various inputs and over regions with different climates. Every model outperforms bicubic interpolation on nearly every performance metric employed throughout this study. Only the 100K timestamp model exhibits higher minumum and maximum prediction errors over eastern Asia, while all other statistics remain competitive. Medians of interpolated images' JS value distributions are between 3 and 5 times larger than those of the best SR-CNNs in every (out of three) regions and at every (out of three) resolutions, indicating much greater discrepancy in the fine-scale SWS prediction distributions. Mean absolute error percentages at various percentiles are significantly higher for interpolated images than SR-CNN predictions in all cases  except the 3-channel 100K model in out-of-sample regions. The SR-CNNs also outperform interpolation in extreme value prediction, particularly for 25K and 48K models. This result indicates the need for adequate input predictors and sophisticated models to recover fine-scale information from coarse one. 
SR-CNN models incorporating coarsened wind speed and high-resolution topography data perform best relative to other models across different resolutions based on a number of performance metrics: JS value, error in test predictions, and ability to capture extreme wind speeds. These model architectures also exhibit some of the smoothest training and validation loss curves compared with other configurations, 
indicating more stable behavior of the optimizer during training.
Aside from the two input fields, the downscaling factor plays the largest role in overall model performance. 
Our SR-CNN framework does not appear to be sensitive to the size of the geographical region of interest. Earlier stages of this study implemented downscaling over a subset of the larger southwestern U.S. region, which exhibited no notable difference in  prediction capability compared with the current domain.

\subsection{Position within existing literature}

In this study we propose a mixture of Weibull distributions to benchmark the SR-CNN. 
Our use of the Weibull mixture arose from \cite{zhang2016quantifying}'s parameterization as a single Weibull pdf to represent the subgrid-scale variability of SWS.   
Fitting a mixture of two Weibull pdfs allows for greater flexibility in quantifying the multimodality of the SGS distribution of SWS. 
This multimodality of the SGS distribution arises from the granularity of  SCREAM data that is finer than the one from the data used by \cite{zhang2016quantifying} and that captures the complex distribution of SWS over various geographies. 
We highlight that \cite{zhang2016quantifying}'s Weibull parameterization of SGS entirely relies on coarse-scale information; in other words,   the Weibull distribution is parameterized with statistics of the coarse data. 
In the present case, because of the multimodality of the SGS distributions, we were able to express the mixture parameters as statistics of the coarse-scale data. Hence we used maximum likelihood to estimate the parameters from fine-scale data and its SGS distributions. 
The MLE-fitted mixtures of Weibull distributions could be used for SGS representation; however, the MLE needs to be performed on fine-scale data at each coarse grid-cell level. 
The capability of the SR-CNNs to output fine-scale SWS with statistical similarity to ground-truth data comparable to the Weibull mixture is indicative of our model's capturing the subgrid variability of wind speed at reduced computational cost.

The SCREAM model provides highly granular 10 m high and 15-minute instantaneous SWS that differs from data used by \cite{stengel2020adversarial} and \cite{kurinchi2021}, which examine downscaling capability over daily averaged wind speeds at higher altitudes. 
Overall, our SR-CNNs exhibit proven efficacy at large downscaling factors. The SR-CNNs operating at a downscaling factor of 8x---the smallest investigated in this study---perform competitively against other downscaling models across a gap of 4--6x (\cite{kurinchi2021}).  
Predictability of 48K and 100K models over the southwest U.S. region, where all models were trained, is comparable to that of the 25K models. We posit that such models can be utilized well over regions that have been used explicitly for training. We also found high predictive capability in unseen regions of lower overall wind speed variability. This is seen in the SR-CNN performance at all resolutions over the eastern Sahara. Eastern Asia performance at  higher downscaling factors is typically lower than that of U.S. predictions, likely because of the larger variability seen in wind speed over this region. 
Our models are unique in that the downscaling operation is performed at one layer, compared with the models of \cite{stengel2020adversarial}, which perform intermediate downscaling operations between coarse inputs and fine output, thus achieving 50x downscaling factor.

\subsection{Perspectives and future directions}
Incorporating  temporal information in our models can be expanded to account for spatial relationships within a given time interval. Future work should consider the inclusion of spatially varying and sub-daily information,  for instance, solar irradiance as an input layer \cite{long95,bett2015}. This feature may be worthy of investigation as a proxy for encoding diurnal patterns seen in wind speed data. \cite{long95} notes that irradiance varies widely based on geographic location, indicating that this feature can provide more useful, fine-grained information rather than an encoded timestamp. 
Additional work may consider the native grid structure of the original wind speed data. In this study we interpolated the climate model data for ease of SR-CNN training, but adapting these models to be compatible with unstructured data will allow for these SR-CNNs to  interact directly with climate model predictions and  undergo further on-line training and refinement, as in the case of \cite{bretherton2022}.  
The Gumbel SR-CNN shows first results in producing nondeterministic downscaled predictions. However, the outputed Gumbel pdfs represent the SWS variability at the 3 km grid-cell level; hence, more work is needed to enforce spatial correlation of the probability model outputs. 

The results of this study highlight the importance of careful feature selection and downscaling factor in the pursuit of accurate statistical representations of downscaled surface wind speeds. Given adequate training time,
state-of-the-art convolutional models are capable of accurate predictions of true subgrid-scale  wind variability across different resolved scales.
By improving on SR-CNN architectures for larger downscaling factors (up to 32x) and secondarily expanding our temporal encoding framework, these models can serve as a marked solution to the computational complexity present in generating realistic, fine-scale climate model information.

\section*{Acknowledgment}
We thank Mihai Anitescu, Adam H. Monahan, and Michael L. Stein for their constructive and helpful comments throughout this work. 
We  thank the U.S. Department of Energy   Science Undergraduate Laboratory Internships (SULI) program for supporting Daniel Getter's initial effort. 
Subsequent efforts of Daniel Getter are supported by the U.S. Department of Energy, Office of Science, Office of Advanced Scientific Computing Research (ASCR) under Contract DE-AC02-06CH11357, and Office of Biological and Environmental Research (BER), under Contract No. DE-AC02-06CH11357.
The effort of Julie Bessac is  supported by the U.S. Department of Energy, Office of Science, Office of Advanced Scientific Computing Research (ASCR) under Contract DE-AC02-06CH11357 and the ASCR Scientific Discovery through Advanced Computing (SciDAC) program through the FASTMath Institute under Contract DE-AC02-06CH11357 at Argonne National Laboratory. 
Y. Feng  acknowledges the support of Argonne National Laboratory  provided by the U.S. DOE Office of Science, under Contract No. DE-AC02-06CH11357, as part of the E3SM project funded by the Office of BER. We thank all the E3SM project team members for their efforts in developing
and supporting the E3SM and SCREAM. 
We gratefully acknowledge the computing resources provided on Bebop and Swing, high-performance computing clusters operated by the Laboratory Computing Resource Center at Argonne. 


\section*{Data} The SCREAM output used in this paper is available as part of the DYAMOND2 intercomparison at \url{https://www.esiwace.eu/services/dyamond}. Code used to conduct model training, validation, and analysis can be found at \url{https://github.com/DanielGetter/Downscaling_SRCNN}.

\bibliographystyle{alpha}
\bibliography{references}

\end{document}